%% file: ms.tex
\def\HA{{H$\alpha$}}

\def\kms{{\:\rm{km\,s^{-1}}}}

\def\FLUXARCSEC{\:{\rm ergs\:cm^{-2}\:s^{-1}\:arcsec^{-2}}}

\def\VEL{\:{\rm km\:s^{-1}}}

\def\feii{\ion{Fe}{2}}
\def\siii{\ion{Si}{2}}
\def\ha{{H$\alpha$}}

\documentclass[12pt,preprint]{aastex}

\begin{document}
\tighten

\newcommand{\MSOL}{\mbox{$\:M_{\sun}$}}  

\newcommand{\EXPN}[2]{\mbox{$#1\times 10^{#2}$}}
\newcommand{\EXPU}[3]{\mbox{\rm $#1 \times 10^{#2} \rm\:#3$}}  
\newcommand{\POW}[2]{\mbox{$\rm10^{#1}\rm\:#2$}}
\newcommand{\SING}[2]{#1$\thinspace \lambda $#2}
\newcommand{\MULT}[2]{#1$\thinspace \lambda \lambda $#2}
\newcommand{\CHINU}{\mbox{$\chi_{\nu}^2$}}
\newcommand{\vsini}{\mbox{$v\:\sin{(i)}$}}
\newcommand\perpix{\:{\rm pixel}^{-1}}
\newcommand\peryr{\:{\rm yr}^{-1}}

\newcommand{\fuse}{{\it FUSE}}
\newcommand{\hst}{{\it HST}}
\newcommand{\iue}{{\it IUE}}
\newcommand{\euve}{{\it EUVE}}
\newcommand{\einstein}{{\it Einstein}}
\newcommand{\rosat}{{\it ROSAT}}
\newcommand{\chandra}{{\it Chandra}}
\newcommand{\xmm}{{\it XMM-Newton}}
\newcommand{\swift}{{\it Swift}}
\newcommand{\asca}{{\it ASCA}}
\newcommand{\galex}{{\it GALEX}}

\newcommand{\msun}{M_\odot}
\newcommand{\src}{G1.9+0.3}
\newcommand{\tbn}{$\theta_{\rm Bn}$}
\newcommand{\roll}{$\nu_{\rm roll}$}
\def\etal{{et~al.}}

\catcode`\@=11
\newcommand{\gapprox}{\mathrel{\mathpalette\@versim>}}
\newcommand{\lapprox}{\mathrel{\mathpalette\@versim<}}
\newcommand{\propapprox}{\mathrel{\mathpalette\@versim\propto}}
\newcommand{\@versim}[2]
  {\lower3.1truept\vbox{\baselineskip0pt\lineskip0.5truept
\ialign{$\m@th#1\hfil##\hfil$\crcr#2\crcr\sim\crcr}}}
\catcode`\@=12


\shorttitle{High-Resolution X-Ray and Optical Study of SN\,1006}
\shortauthors{Winkler \etal}

\slugcomment{Accepted for publication in {\it The Astrophysical Journal}}

{\title{A High-Resolution X-Ray and Optical Study of SN\,1006: Asymmetric Expansion and Small-Scale Structure in a Type Ia Supernova Remnant\footnote
{Based on observations made with NASA's {\em Chandra X-ray Observatory},  operated by the Smithsonian Astrophysical Observatory 
under contract \# NAS83060;   the data were obtained through program GO1-12115.}}

\author{
P. Frank Winkler,\altaffilmark{2}
Brian J. Williams,\altaffilmark{3} 
Stephen P. Reynolds\altaffilmark{4}
Robert Petre,\altaffilmark{3} 
Knox S. Long,\altaffilmark{5}
Satoru Katsuda,\altaffilmark{6} and
Una Hwang\altaffilmark{3}
}
\altaffiltext{2}{Department of Physics, Middlebury College, Middlebury, VT, 05753; winkler@middlebury.edu}
\altaffiltext{3}{NASA Goddard Space Flight Center, Greenbelt, MD 20771; brian.j.williams@nasa.gov; robert.petre-1@nasa.gov}
\altaffiltext{4}{Physics Department, North Carolina State University, Raleigh, NC 27695; reynolds@ncsu.edu}
\altaffiltext{5}{Space Telescope Science Institute, 3700 San Martin Drive, Baltimore, MD, 21218;  long@stsci.edu} 
\altaffiltext{6}{RIKEN (The Institute of Physical and Chemical Research), 2-1 Hirosawa, Wako, Saitama 351-0198, Japan}

\begin{abstract}

We introduce a deep (670 ks) X-ray survey of the entire SN\,1006 remnant from the {\em Chandra X-ray Observatory}, together with a  deep  \ha\  image of SN\,1006 from the 4m Blanco telescope at CTIO.  Comparison with {\em Chandra} images from 2003 gives  the first measurement of the X-ray proper motions around the entire periphery, carried out over a nine-year baseline.  We find that 
the expansion velocity varies  significantly with azimuth.  The highest velocity of $\sim 7400\kms$ (almost 2.5 times that in the NW) is found along the SE periphery,  where both the kinematics and the spectra indicate that most of the X-ray emission  stems from ejecta that have been  decelerated little, if at all.  Asymmetries in the distribution of ejecta are seen on a variety of spatial scales.  
Si-rich ejecta are especially prominent in the SE quadrant, while O and Mg are more uniformly distributed, indicating large-scale asymmetries arising from the explosion itself.  Neon emission is strongest in a sharp filament just behind the primary shock along the NW rim, where the pre-shock density is highest.  Here the Ne is likely interstellar, while Ne within the shell may include a contribution from ejecta.  Within the interior of the projected shell we find a few isolated ``bullets" of what appear to be supernova ejecta that are immediately preceded by bowshocks seen in \ha, features that we interpret as ejecta knots that have reached relatively dense regions of the surrounding interstellar medium, but that appear in the interior in projection.  Recent three-dimensional hydrodynamic models for Type Ia supernovae display small-scale features that strongly resemble the ones seen in X-rays in SN\,1006; an origin in the explosion itself or from subsequent hydrodynamic instabilities both remain viable options.  We have expanded the search for precursor X-ray emission ahead of a synchrotron-dominated shock front, as expected from diffusive shock acceleration theory,  to numerous regions along both the NE and SW rims of the shell.  Our data require that a precursor be thinner than about 3\arcsec, and fainter than about 5\%\ of the post-shock peak.  These limits suggest that the magnetic field  is amplified by a factor of 7 or more in a narrow precursor region, promoting diffusive particle acceleration.

\end{abstract}


\section{Introduction \label{sec_intro}}
\label{intro}

X-ray investigation of the SN\,1006 remnant began with the suggestion that a bump on one edge of the extended Lupus Loop supernova remnant, observed during a 1971 rocket flight by the Livermore group, might be attributed to SN\,1006 \citep{palmieri72}.  Positive identification followed from the X-ray experiment on the OSO-7 satellite \citep{winkler76}, and it was included in the final {\em Uhuru} catalog \citep{forman78}.  Its flux, about 3\% of that from Cas A and $\lesssim 0.2\%$ of that from the Crab Nebula, made it among the faintest sources identified by the first generation of X-ray astronomy satellites.   

After that humble beginning, SN\,1006 was studied with virtually all  major X-ray satellites through the next two decades:  {\em SAS-3} \citep{winkler79}, the {\em Einstein Observatory} \citep{pye81}, \rosat\ \citep{willingale96, winkler97}, and others.  The seeming discord between a remnant with a clear bilateral shell  in radio maps, yet with a featureless power-law spectrum in its bright X-ray regions, led to a well-known paper from the {\em Einstein}/{\em OSO-8} era, ``Is the remnant of SN\,1006 Crab-like?"  \citep{becker80}.  The definitive answer came from \citet{koyama95}, whose spatially resolved spectroscopy using \asca\ showed that the X-ray-bright NE and SW shell limbs are indeed power-law dominated, but that emission from the interior and the remaining shell is dramatically different:  soft, thermal X-rays more typical of supernova remnants (SNRs).  This finding provided the  first clear evidence for diffusive shock acceleration of  electrons to high energies in SNR shocks, and cemented the long-suspected link between supernovae and cosmic rays.

SN~1006 is relatively close at 2.2 kpc \citep{winkler03}, where the spatial scale is $1\arcsec \approx 0.01$ pc, and since it is located 14.6\degr\ above the Galactic plane it has relatively low foreground absorption, $N_H \approx 7 \times 10^{20}\;{\rm cm}^{-2}$ \citep{dubner02, uchida13}.\footnote{\citet{nikolic13} have recently argued for a shorter distance of 1.7 kpc to SN~1006, based on new optical spectra and  model calculations from \citet{vanadelsberg08}. We use the 2.2 kpc distance throughout this paper; scaling to a shorter distance is straightforward.}  Its location, low-density surroundings far from any recent star formation, apparent absence of any compact remnant, and the implication from Chinese records that it remained visible for several years \citep{stephenson02} all indicate that it was a Type Ia event; it is the closest remnant of a historical SN\,Ia.  

All these attributes have made it an important target for continuing investigation from the current generation of X-ray telescopes: {\em Chandra}, {\em XMM-Newton}, and {\em Suzaku}.  The first {\em Chandra} observations were a pair of deep pointings in 2000 and 2001 with the Advanced CCD Imaging Spectrometer (ACIS) S-array along the contrasting nonthermal NE and thermal NW rims in 2001, reported by \citet{long03},  followed in 2003 by a mosaic of ACIS-I fields that covered the entire remnant \citep{cassam-chenai08}, and in 2008 by a second-epoch ACIS-S observation of the NE rim \citep{katsuda09}.   The {\em Chandra} observations we report here were designed as a follow-up to those by \citet{cassam-chenai08} with nearly five times longer exposure around most of the rim, in order to provide both a more detailed look at small-scale features and a second epoch for measuring the expansion.

From {\em XMM-Newton}, a very deep set of imaging observations has been presented by \citet{miceli12, miceli13b} and references therein, while   observations with the reflection grating spectrometer have focused on a prominent knot of ejecta along the NW shock front \citep{vink03, broersen13}.  Finally, {\em Suzaku} results focusing on spatially-resolved spectroscopy have been reported by \citet{yamaguchi08}, \citet{bamba08}, and \citet{uchida13}.  
Results from these missions that are most relevant to the  present paper will be mentioned in the context of the new {\em Chandra} data in subsequent sections.

In this paper we present the first  results from the complete  \chandra\ Cycle 13 Large Project Observation of SN~1006, comprising pointings at 10 overlapping fields with the ACIS for a total of 670 ks.  We also present the images resulting from a deep optical study from the 4m Blanco telescope and Mosaic II camera at CTIO, carried out in 2010 April.  Conceived together, these are intended to give a detailed high-resolution view of this important remnant in multiple bands, to which we plan to add  high-resolution radio images to be carried out from the Jansky Very Large Array in its three hybrid configurations over the next two years.  In Sections \ref{xray_obs} and \ref{optical_obs} of this paper we briefly describe the X-ray and optical observations and data reduction, and present the complete mosaic images.

We then highlight some results that are immediately apparent, and suggest areas for future work by ourselves and others.  In Section \ref{features} we discuss some of the X-ray and optical features that the new images reveal, including what appear to be ``bullets" of ejecta preceded by small optical bowshocks.  Section \ref{proper_motion} presents the first  X-ray expansion measurement around the entire rim of the 30\arcmin\ diameter shell;  Section \ref{sec:ejecta} presents   narrow-band X-ray images in lines corresponding to different important elements, with inferences  for the distribution of ejecta; and Section \ref{sec:halo} presents limits on precursor X-ray emission ahead of the synchrotron-dominated shocks.   We present a  discussion of these results and implications for SN~1006 in particular, and for Type Ia SNe in general, in Section \ref{sec:discussion}.   Finally, Section \ref{sec:conclusions} summarizes our conclusions from this paper.   


\section{X-ray Observations from the {\em Chandra} ACIS\label{xray_obs}}

The goal of the  new  \chandra\  observations was to provide a detailed picture of  the fine-scale structure of the X-ray remnant and to provide a second-epoch image in order to measure the expansion around the entire shell.    A series of deep ACIS pointings was planned to cover the entire 30\arcmin\ diameter remnant, with most of the aim points  located just inside the shell rim,  spaced to ensure coverage of the entire rim close enough to on-axis to achieve a resolution $\lesssim 5\arcsec$. 
The S-array was used for the NE and NW rims  (ObsIDs 9107 in 2009 and ObsID 13737 in 2012, respectively) in order to match earlier ACIS-S pointings for proper-motion measurements along those parts of the shell \citep{long03}.  Results of both these measurements have already been reported by \citet{katsuda09,katsuda10b, katsuda13}.   

All the other pointings were made with the ACIS-I array and were carried out in 2012 April-July.  The strategy was similar to that used by \citet{cassam-chenai08} for their ACIS-I mosaic, except that most of our pointings were 80-100 ks instead of 20 ks, and ours emphasized positions near the shell rim in order to achieve optimum angular resolution there.   We also benefited from their earlier image in being able to position the detector to avoid having critical features fall onto chip gaps in the ACIS detector.  Despite the deep second-epoch observation of the NE rim in 2008, we included short observations of this bright region in 2012 in order to ensure contemporary observations of the entire shell.    
A complete journal of the new observations is provided in Table  \ref{tab:acis_obsns}, and a map showing the relative exposure is in Figure \ref{fig:xray_expmap}.  All the data were processed through the standard \chandra\ pipeline, and then reprocessed using CIAO version 4.5 and CALDB 4.5.6, in order to assure that the latest gain and quantum-efficiency corrections are used.  

\subsection{Aspect Correction and X-ray Mosaics}

In order to correct the data  to a uniform, absolute coordinate system that can be used for precise comparison with optical and other data, we compiled a list of astrometric optical sources from the NOMAD catalog \citep{zacharias05}, selected for position errors $< 500$ mas and proper motions $< 50\: {\rm mas\: yr}^{-1}$ in both R.A. and Decl.   For each ObsID we used the CIAO task {\tt wavdetect} to locate point sources in the field, and from these selected only ones  with significance $> 5\,\sigma$.  We then used the CIAO task {\tt wcs\_match} to match the source list for each field against our selected list from the NOMAD catalog, and to calculate translations to achieve the optimum match.\footnote{The {\tt wcs\_match} task has an option to calculate transformations that include rotations and scale changes, but we used the translation-only method, which is more stable for a small number of sources.}  For most of the X-ray pointings there were 4 to 6 excellent source matches that yielded aspect translations of $<1.4$ ACIS pixels $(< 0\farcs 7)$ in both coordinate directions.   For three pointings, however (ObsIDs 13738, 13743. and 14423), there were not enough reliable source matches, so no aspect correction was applied.   Comparison of the {\tt wavdetect} positions for point X-ray sources detected with high significance in overlapping ObsIDs ({\em not} limited to sources with optical counterparts) showed  agreement to within $< 1\  {\rm ACIS\ pixel} \approx 0\farcs 5$ rms. 

We then combined all the aspect-corrected and reprocessed 2012 data into  mosaic images of SN~1006 in several energy bands using the CIAO script {\tt merge\_obs}.   The resulting mosaic image in the soft (0.5-1.2 keV, shown in red), medium (1.2-2.0 keV, shown in green), and hard (2.0-7.0 keV, shown in blue) bands is shown in Figure \ref{fig:xray_rgb}.

\section{Optical Images from CTIO\label{optical_obs}}

The most prominent optical emission from SN~1006 is a  relatively bright, delicate filament extending along much of the NW limb of the shell \citep{vandenbergh76, long88, raymond07} and seen only in the Balmer lines of Hydrogen \citep{schweizer78, ghavamian02}.   In addition, \citet{winkler03} reported far fainter and less distinct Balmer emission surrounding  most of the shell, and there is an even fainter diffuse oval of emission, probably associated with SN~1006, filling much of the northern half of the remnant.  The  ``nonradiative" emission like that from SN~1006 results from at least partially neutral pre-shock H atoms that traverse the shock, where they can undergo either direct collisional excitation, or charge exchange with hot post-shock protons---processes that produce narrow and broad emission-line components, respectively \citep[][for a recent review]{chevalier78, heng10}.  

In order to study the emission from the entire remnant in greater detail,  we carried out  deep optical imaging of SN~1006 in two narrow bands: \ha\ and  a matched  continuum for subtracting the stars to reveal the faintest Balmer emission  features.  
The observations, in 2010 April, used  the 4m Blanco telescope at CTIO and Mosaic II camera, whose field of $37\arcmin \times 37\arcmin $,  at a scale of  $0\farcs 27 \perpix$, is well matched to the size of the remnant.  The filters were centered at 6563 \AA\ and 6650 \AA, respectively, both with a bandwidth of 80 \AA\ (FWHM). We obtained 24 10-min exposures in \ha, and 22 10-min ones in the continuum, dithered by a few arcmin between exposures to cover a somewhat larger total field and to improve the flat-fielding.  The observational details are summarized in Table \ref{tab:ctio_obsns}.   

The images were processed through the standard NOAO Mosaic pipeline; subsequently we determined a more precise World Coordinate System using stars from the UCAC4 catalog \citep{zacharias13}, selected for position errors $< 100\;$mas and proper motions $< 10\; {\rm mas} \peryr$.  With typically 80-100 stars on each of the 8 chips in each Mosaic frame, we obtained excellent fits, with rms uncertainty that was typically $\lesssim 60$ mas in both R.A. and Decl.   Both the \ha\ and continuum frames were then reprojected onto a standard system at a scale of $0\farcs2 \perpix$, and stacked to produce mosaic images, using IRAF\footnote{IRAF is distributed by the National Optical Astronomy Observatory, which is operated by AURA, Inc., under cooperative agreement with the National Science Foundation.} tasks in the {\tt mscred} package.  Finally, we scaled and subtracted the continuum image from the \ha\ one, to give the image shown in Figure \ref{fig:ha_image}.   

The basic  morphology of optical  emission from SN~1006 is, naturally, entirely consistent with previous observations.    
The brightest segments of the NW filament have \ha\ surface brightness 3.5 to $4 \times 10^{-16}\;\FLUXARCSEC$.  In our earlier deep image, taken in 1998 from the  0.6/0.9 m Curtis Schmidt telescope at CTIO, we measured about half this surface brightness \citep{winkler03}; since the sharpest and brightest segments were blurred in the lower-resolution Schmidt images, the values from the two observations are entirely consistent.   
The far fainter and more diffuse parts of   the  rim that are clearly visible in the S and elsewhere have surface brightness $\sim 1 \times 10^{-17}\;\FLUXARCSEC$, fainter by a factor of $\sim 40$ than the brightest ones.  Our 1998 Schmidt image shows these features as well, at about the same surface brightness but with  lower   signal-to-noise \citep[Figure 5 of][]{winkler03}.   For reference, we show both the 1998 and 2010 continuum-subtracted images at the same scale in the lower two panels of Figure~\ref{fig:ha_image}.   The images from both epochs show additional faint structures; e.g., the previously mentioned  faint, diffuse oval of emission that fills much of the northern half of the shell, and an irregular band of emission that wraps around the southern half of the  shell, about 5\arcmin\ inside the rim.   The latter connects to  emission that extends far beyond the shell to the west and north, and that is presumably in the foreground or background and not physically associated with SN~1006 itself.   It is not clear just which emission features within the shell are physically associated, but some---ones with associated X-ray features---definitely are, as we discuss in the following section.   

\section{Relation of X-ray and Optical Features \label{features}}

Comparison between the X-ray and optical images shows several thin arcs of Balmer emission, primarily within the  southern portion of the SN~1006 shell, that lie immediately in front of some of the brighter tufts or flocculi  of X-ray emission.   These X-ray structures, also seen in  previous X-ray images, have scales that are  typically 10\arcsec - 30\arcsec\ (0.1 - 0.3 pc).   Two examples are shown in Figure \ref{fig:bowshocks}.  These  Balmer filaments seen (in projection) in the remnant interior strongly resemble bowshocks, and the X-ray tufts behind them have spectra indicating that they are ejecta-dominated (see Section \ref{sec:ejecta}).  These are probably  similar structures to the far brighter bulge in the NW Balmer filament, at about 2 o'clock in Figure 3, which precedes a bright thermal X-ray knot that has long been attributed to an ejecta bullet \citep{long03, vink03, broersen13}.   

The presence of  Balmer emission absolutely requires partially neutral interstellar H ahead of the shock, so the bowshock features must be located on the front or back sides of the remnant's shell, seen in the interior only in projection.  The X-ray knots behind them have a somewhat flattened appearance, consistent with ejecta running into interstellar material.   
There are many small X-ray tufts similar to those shown in Figure \ref{fig:bowshocks}, and with a spectral character that indicates SN ejecta, but that are {\em not} preceded by optical bowshocks.   This absence simply indicates the absence of neutral gas in front of them; they may not have reached the remnant shell, or the pre-shock gas at that point could be fully ionized or too tenuous to produce significant Balmer emission.
The origin of the X-ray tufts---whether with or  without associated Balmer bowshocks---is not obvious; they could have resulted from small-scale density inhomogeneities imprinted during  the explosion itself \citep{orlando12}, or they could be the result of more recent Rayleigh-Taylor instabilities in the expanding ejecta \citep[e.g.,][]{warren13}.   We discuss these possibilities further in Section \ref{disc:small}.   There are also several thin arcs of Balmer emission without an obvious X-ray knot behind, which could have resulted from less dense clumps of ejecta or ones that have dissipated.

In the NW, the new \ha\ image clearly shows the  complex structure {\em ahead of} the bright filaments, best shown in Figure \ref{fig:nw_shock} (center),  where this region is displayed with a very hard stretch to show the faintest emission.  Very faint X-ray emission is also seen outside the main Balmer filament, up to the outermost limit of optical emission.  The optical morphology  indicates a rippled sheet seen edge-on, with the multiple edges representing tangencies at different locations \citep[as shown by][]{hester87}.  It has long been clear that this is the cause for the undulating structure of   the primary NW filament, but the deeper image shows this structure to be  more complex than previously realized.  The bright filament is the result of an encounter between the primary SN shock and a denser ambient medium than  around most of SN~1006, nd produces the rather flattened structure and slower expansion than elsewhere \citep{katsuda13}, as well as the only IR emission seen in SN~1006 \citep[24$\,\mu$m emission just behind the Balmer filament,][]{winkler13}.  However, the more complex structure indicates that the dense region that has led to the bright Balmer, X-ray, and IR emission in the NW must not be  a ``wall," but is instead limited in extent along the line of sight, so that the primary shock has passed well beyond it on either the front or back side of the shell.

\section{X-Ray Proper-Motion Measurements\label{proper_motion}}
A proper-motion measurement for the bright, synchrotron-dominated E-NE rim was reported by \citet{katsuda09}, and a similar measurement for the thermal-dominated NW,  using an observation from the current project as the second epoch, by \citet{katsuda13}.   Both these measurements compared many individual features in comparably deep ACIS-S observations:  epochs 2000-2008 for the E-NE, and 2001-2012 for the NW\@.   The results showed that in the NW, the brightest X-ray filaments that lie just within the bright Balmer filaments have a velocity of $\sim 3000 \kms$, essentially the same as measured by \citet{winkler03} for the near-coincident Balmer filaments, but that two fainter knots in the NW have nonthermal spectra and much higher  velocities: $\sim 5000 \kms$, essentially the same as measured for the synchrotron-dominated filaments in the NE.\footnote{As previously mentioned, throughout this paper we assume a distance of 2.2 kpc   for converting from proper motion to shock velocity, in part to facilitate comparison with previous proper-motion studies.}

To study the expansion around the {\em entire} shell, we have  used as the first-epoch image that obtained from a set of eleven overlapping ACIS-I exposures from 2003 \citep[J. Hughes, PI,][]{cassam-chenai08}, each with an exposure time of about 20 ks.  We reprocessed this data set using the same CIAO and CALDB versions as for the 2012 data.  In order to assure that the data from the two epochs were accurately registered, we began with the set of X-ray point sources located by {\tt wavdetect} in each of the ten 2012 fields (after our small aspect corrections), and selected only those with significance $> 6\sigma$.   For the many duplicate sources (the result of field overlap), we kept only the one with the smallest error ellipse  in each case,  to give a master list of 129 distinct sources.   We then determined the small aspect correction for each of the eleven fields from 2003 by  using {\tt wavdetect} on each field and matching the resulting source list  against our master list.  (This procedure is preferable to registering the 2003 data using the NOMAD optical catalog---as with the 2012 data---because there are far more X-ray point sources than astrometric stars, and because fewer of those stars were detected with high significance in the shorter 2003 X-ray exposures.)   Finally, we used {\tt merge\_obs} to combine the 2003 data, just as with that from 2012.   

The structure of SN~1006 appears virtually identical in both data sets, though the deeper 2012 data reveal it in greater detail.  By aligning and blinking the 2003 and 2012 images, however,  expansion of the shell becomes  obvious.  The expansion is shown somewhat less dramatically in Figure \ref{fig:xray_diff}, which is simply the difference between the merged images at the two epochs.  The expansion is most obvious in the NE and SW, where the shock front is most sharply defined and the X-ray emission is brightest.  While the expansion of other regions around the shell is less obvious in the difference image, it is clear on blinking the images, as in the animation that appears in the on-line version of Figure \ref{fig:xray_diff}. 

In Figure~\ref{fig:azimuth_plot} we show the measured proper motion   as a function of azimuthal angle (measured in the conventional sense, rotating eastward from north).  
We took the center to be at R.A. $= 15^{\rm h} 02^{\rm m} 54\fs 9$, Decl.\ $= -41\degr 56\arcmin 08\farcs9$ (J2000.), the same as that defined by \citet{katsuda09}, and extracted radial profiles in 10\degr\ azimuthal sectors from the merged, aspect-corrected images for both 2003 and 2012.   We then carried out a minimum-$\chi^2$ analysis {\em limited to the outermost edge of clear X-ray emission} to determine the shifts---in the purely radial direction, regardless of the orientation of individual features---between the two epochs, which are separated in time by 9.1 years.  (The 2003 observations were all carried out over 4 days, 2003 April 8-11.  Those in 2012 were spaced over two months, from 2012 April 20 to 2012 June 15; we  used the mean epoch.)   The procedure is similar to that we used in earlier analyses of the NE and NW rims \citep{katsuda09, katsuda13}, except in those papers we identified and measured the proper motion for individual identifiable features, in a direction normal to the local shock surface, which in some cases was significantly nonradial.\footnote{Another significant difference is that in both the NE and NW measurements, the first-epoch images were far deeper than the 2003 one used here.}  
In order to provide consistency around the entire circumference, including those regions with  no crisply defined features, in  the present measurement we  simply used the portion of the periphery that falls within an azimuthal sector, and  measured the purely radial motion.  We show  typical profiles from both epochs, one from each of the four quadrants, in Figure~\ref{fig:pm_profiles}.\footnote{In the NW, we made an exception to measuring the outermost edge of X-ray emission:  There the very faint emission beyond the bright filament, described in the previous section, is simply to faint to give a measurement.  Instead, the measurement at azimuths about 310\degr\ to 350\degr\ is really for the bright (thermal) X-ray filament.  An example is shown in Figure~\ref{fig:pm_profiles}, upper right.}

For the synchrotron-dominated regions in the NE and SW, where the shock is  clearly defined in X-rays, the above method gives  precise measurements, and for most of the NE these are in excellent agreement with those of \citet[][also shown in Figure~\ref{fig:azimuth_plot}]{katsuda09}.   
Only in the 20\degr-30\degr\ sector is there a seeming disagreement, but examination of Figure~1 of that paper shows that the individual features measured by \citet{katsuda09} in the 20\degr-30\degr\ sector, while the brightest at this azimuth, do {\em not} lie  at the outer edge, and are oriented far from normal to the radial direction.  Thus, there is really no disagreement with earlier measurements in the NE.

In most of the thermal-dominated SE quadrant, and in part of the NW, the shock front is not well-defined in X-rays.  Three sectors in the SE (140\degr-170\degr), where the signal-to-noise was low and the individual measurements highly uncertain, were combined into a single 30\degr\ sector.  
In the NW, comparison with the results from \citet[][also shown in the figure]{katsuda13} shows that some of these sectors include parts of both thermal and nonthermal features with quite different shock velocities, so the fact that profiles for the entire sector at different epochs did not correspond closely is hardly surprising.   For two sectors, centered at azimuths 125\degr\ and 295\degr, there was not a sufficiently sharp X-ray limb to yield a proper-motion measurement at all.  All of the measurements shown in Figure~\ref{fig:azimuth_plot} gave satisfactory  fits (reduced $\chi^2 \sim 1$), and there were no systematic differences  in $\chi^2$ as a function of azimuth, post-shock brightness, or measured  velocity.  In addition, we examined each fit by eye to ensure that all look reasonable.  

The most notable fact about the proper-motion measurements is that the expansion velocity in the SE is higher than anywhere else around the shell: $\sim 7400\pm800 \kms$, almost 2.5 times faster than that of the far brighter thermal X-ray filament in the NW.  
In the SE,  the shock front is not really defined at all in X-rays (Figure~\ref{fig:xray_rgb});  instead the outermost emission is marked by tufts that we  interpret as SN ejecta based both on their kinematics and  their spectra (see Section~\ref{sec:ejecta}).  Furthermore, some of these tufts are located beyond the outermost of the multiple indistinct shells seen in \ha\ (Figure~\ref{fig:ha_image}).  

A common way of expressing the proper motion in an SNR is through the expansion parameter $m$: the power-law index in $R \propto t^m$, where $t$ is the age of the remnant.  This parameter can be interpreted as the ratio of current expansion rate divided by the mean rate over the remnant's lifetime, $m=\mu t/\theta$, where $\theta$ is the angular radius.   The outer tufts of emission in the SW are located $\sim 14\farcm8 = 9.5$ pc from the center, so with an age of 1001.5 yr (the mean for the 2003 and 2012 epochs) we find $m = 0.80 \pm 0.08$, close to the free-expansion value of 1.  
This value contrasts sharply with that of $m = 0.54 \pm 0.05$ measured by \citet{katsuda09} for the nonthermal NE shell, which suggested that in the NE SN~1006 is transitioning to the adiabatic phase ($m \sim 0.4$).
It is entirely consistent to interpret the tufts that define the SE X-ray periphery of SN\,1006 as plumes of ejecta that have been coasting almost undecelerated into a very low-density region of the interstellar medium (ISM).   The proper motions we have measured for these tufts do {\em not} represent a shock velocity, but rather the current motions for these ejecta tufts.  

Simulations of either ejecta bullets originating in the SN explosion \citep[e.g.,][]{orlando12} or
Rayleigh-Taylor (R-T) ``mushroom caps''  that formed more recently from hydrodynamic
instabilities \citep[e.g.,][]{warren13} show that denser regions
can move faster than the blast wave. Ejecta ``bullets'' overtake the
blast wave, move beyond its mean radius briefly (as in knots D and E
in Figure~13), and are shredded and dissipate.   R-T ``mushrooms''
also represent regions of denser-than-average ejecta, which formed much
later but which  can also, for highly compressive shocks, penetrate the forward
shock before they die back.  It is,  therefore, not surprising that the
ejecta plumes in the SE currently show higher velocities than the
average blast wave around most of SN~1006.  

Velocities around most of the synchrotron-dominated NE and SW limbs center around
$5000\kms$.  (Exceptions include the ``bulge'' to the north, azimuths $\sim 10\degr$-\,30\degr, that shows a currently
higher velocity than adjacent regions, and a point at about 260\degr\ that may reflect the ``bulge'' apparent in the WSW.)   
Variations in the
velocities are a bit larger in percentage terms than variations in the
remnant radius, implying that the upstream ISM density varies both azimuthally around the original SN location and with distance
from it, as is most clearly demonstrated in the NW.

\section{X-Ray Spectra: Thermal Emission from  Supernova Ejecta and the ISM\label{sec:ejecta}}

In order to investigate the spatial distribution of emission from different elements stemming from SN ejecta or from the ISM, we have produced  equivalent-width (EQW) images in characteristic lines, according to the following  procedure, similar to that introduced by \citet{hwang00} for Cas A.  We first extracted images, binned by a factor of 8 (4\arcsec\ pixels) in  a number of narrow energy bands corresponding to K-line emission from significant elements: O, Ne, Mg, and Si, and also in narrow line-free continuum bands to either side of each of the line image (all the bands are detailed in Table~\ref{eqwtable}).   We then smoothed all the images slightly with a 2-pixel Gaussian filter,  divided each continuum image by its bandwidth in keV, logarithmically interpolated between high and low continuum bands, and subtracted the appropriate continuum from each emission-line image.  Finally, to better distinguish between composition and density effects, we divided each continuum-subtracted image by the appropriate continuum one to produce the EQW images.  These are analogous to optical EQW images, except that in our case  the units are keV instead of \AA.

The resulting images are shown in Figure \ref{fig:eqw}.   In all four images the NE and SW limbs appear dark, indicating little line emission, since  strong synchrotron radiation there dominates any thermal emission.
Within the interior, however, there are distinct differences.  Silicon, expected to stem primarily from the ejecta in a Type Ia SN, is strongly concentrated in the SE quadrant---suggesting a clear asymmetry in either the distribution of Si ejecta or in the (presumably reverse) shock pattern that has heated it.  This confirms the recent {\em Suzaku} result from \citet{uchida13}.  Oxygen and Magnesium show less extreme concentrations in the SE, and also concentrations well inside the shell rim to the NW.   These too probably arise largely from SN ejecta, with significant contributions from the shocked ISM.   Oxygen in particular is also strong  behind the primary shock to the NW.\footnote{Also  important in the ejecta from SN\,Ia is Fe, whose K-lines at 6.7 keV are shown clearly by \citet{uchida13}.     However, the ACIS sensitivity above 5 keV is too low to enable significant measurements in the faint thermal plasma of SN~1006.}

Neon is also prominent in the NW, but strongly concentrated in a narrow filament immediately behind the shock front (Figure \ref{fig:eqw}).  Given the morphology of the Ne filament and its location where both the pre-shock density is highest \citep[e.g.][and references therein]{winkler13} and the overall thermal emission is strongest (Figure \ref{fig:xray_rgb}), it seems certain that this feature arises from shocked ISM.   
Elsewhere within SN~1006, however, the Ne emission is likely a mixture of shocked ISM and ejecta.  The distribution of Ne within the shell most closely resembles that of Mg, which, like Ne, results from Carbon-burning.  One curious feature in the Ne distribution is the relatively strong band curving from about 9 o'clock to 5 o'clock across the SE quadrant.   Both Mg and Si emission are relatively {\em weak} along this same band; the cause of these effects remains under investigation.

In order to further investigate the contrasts that are evident in Figure \ref{fig:eqw}, we have selected the brightest 25\% of the pixels in each of the EQW images, and extracted a combined spectrum from these, with the results shown in Figure~\ref{fig:spectra}. These regions are not mutually exclusive; e.g., the brightest 25\% of the Si pixels include some of the brightest 25\% of the Mg pixels, etc.  A few interesting trends can be seen in comparing these spectra. While all of the spectra clearly contain significant emission from ejecta, the ``Ne-pixel" spectrum is heavily weighted by pixels along the bright NW shock, and hence should emphasize emission from forward-shocked ISM.   \citep[While some ejecta may contribute, especially from Ne-bright pixels in the interior, nucleosynthetic models, e.g.,][predict far less Ne than either O or Si, and somewhat less than Mg, in SN~Ia ejecta.]{nomoto84,iwamoto99,maeda10} 
And indeed this spectrum, in addition to showing a Ne~IX  0.92 keV line that is much stronger (relative to the lines from other elements) than the others in Figure~\ref{fig:spectra}, also shows  O lines with significantly stronger He-like (0.57 keV) than H-like (0.65 keV)  ions.   These O lines indicate that  the ionization state of the gas is lower than in the ejecta-dominated regions, consistent with a spectrum dominated by recently shocked ISM. 

To better understand the spectra resulting from these ``brightest
pixels" in the EQW images, we have modeled each of the spectra with an
identical model in XSPEC\@. We used a single absorbed, variable-abundance
NEI model ({\tt phabs*vnei}), with the absorption column frozen for all
spectra to a value of $7 \times 10^{20}\; {\rm cm}^{-2}$. The temperature and
ionization state of the plasma were allowed to vary, as were the
abundances of the four elements represented in Figure~\ref{fig:spectra}:  O, Ne, Mg, and
Si.   We fit the spectra only up to 2 keV. We stress here that these models are {\em
not} intended to be ``physical" models that accurately represent the
current conditions of the plasma. Such a model would be highly complex,
since each of these four spectra represent an amalgam of spectra from
physical locations all over the remnant, and all undoubtedly contain
emission from both forward-shocked ISM and reverse-shocked ejecta.  (This complexity is demonstrated by the multiple components required to fit the {\em Suzaku} spectra \citep{yamaguchi08,uchida13}.)
Rather, we have simply extracted qualitative differences between the spectra,
primarily in terms of the relative abundances implied by the model fits.\footnote{Other effects may play a role as well:  the 0.74-0.87 keV band, which we use as a continuum for both the O and Ne EQW images, includes a ``false continuum" produced by blended Fe-L lines, so an anomalously  low Fe abundance may artificially increase the O and Ne EQW.    A similar effect for the O images only may be produced from anomalous N abundance, since the low continuum for the O EQW images includes lines from N.  Finally, temperature variations can also affect the EQW values.  Detailed discussion of these issues is beyond the scope of this initial report.} 

The model fits confirm that the EQW images are indeed showing meaningful variations in the abundances of the respective metals.  For instance, the ratio of Si/Ne is approximately twice as high in the Si-selected spectrum of Figure~\ref{fig:spectra} compared to the Ne one, and the O/Ne ratio is 70\% higher in the O-selected spectrum compared to the Ne one---with small formal errors in both cases.\footnote{All abundance ratios are the abundances by number, relative to Solar.}   Our spectra show that while the broad-band image of SN~1006 may appear relatively uniform in the interior, one can still use regions of strong Si and S, at least, to identify regions dominated by  ejecta. 

Finally, we have extracted a combined spectrum from the two ejecta ``bullets"  preceded by \ha\ bowshocks shown in Figure~\ref{fig:bowshocks},   subtracted a local background from several  diffuse regions nearby, and have fit this with the same {\tt phabs*vnei} model in XSPEC.   The result requires strong over-abundances of Si ($\sim 6\times$ solar) and S ($\sim 11\times$ solar), and an under-abundance of Ne ($\sim 0.2\times$ solar).  The paucity of  counts in these spectra, together with the problem of the bullets being superimposed on the more diffuse background,  limit the quantitative validity of the fits, but there can be little doubt that they are, indeed, composed primarily of  ejecta.

\section{A Shock Precursor?}\label{sec:halo}

A firm prediction of diffusive shock acceleration theory is that
accelerated electrons will spend some of their time ahead of the
shock, producing synchrotron radiation in a ``halo'' of X-ray
emission.  A precursor X-ray halo has yet to be conclusively
identified in any young SNR, but SN~1006, which is nearby, with low
foreground absorption and well-defined synchrotron rims, probably
presents the best opportunity to detect one.  
As we discuss below, two potential observables for a precursor halo
are the spatial extent and the magnitude of the sharp jump at the
shock front.

\citet{long03} used 
 {\it Chandra} observations from 2001 to search for halo emission in the NE region, and found  that any halo is very faint, with a mean surface brightness $\lesssim 1.5\%$ of
the peak surface brightness at the shock front, or is very thin, limited in extent to under a few arcsec ahead of the shock.  
Subsequently, \citet{morlino10} described a model in which preshock emission rises steeply
prior to the shock over a scale of $\sim 10\arcsec$, followed by a final jump by an
order of magnitude to reach the peak emission---due to a magnetic-field jump
of a factor of $\sim 4$ at the ``viscous subshock'' (see below).  They found this 
model halo to be consistent with profiles observed in the NE by \citet{long03}, and  also with the 2008 epoch observation of the same region \citep{katsuda09}.
 
\subsection{New Observational Results}
  
Here, we report a similar analysis to that of \citet{long03}, but carried out in greater detail  over a
larger sample of regions of interest. We have selected six regions
shown in Figure~\ref{fig:halo_regions}: four  along the NE limb and two along the   SW.
All of the regions are located in places where the local shock
front is nearly linear, and are 20\arcsec\ to 40\arcsec\ wide
(depending on the length of a clean segment of the rim), oriented
perpendicular to the front.  \citep[Our region E-1 is  similar to the one chosen by][]{morlino10}.
In each case we selected the longest single observation for
which each region was closest to on-axis, replaced any obvious point
sources with local background, and then extracted profiles
perpendicular to the shock front, to give the results shown in 
Figure~\ref{fig:halo_profiles}.   We used  unbinned
exposure-corrected flux images from 1-4 keV, eschewing both low- and
high-energy emission to minimize background contamination.   For each profile in the figure we also show the response expected from a sharp edge of emission, folded through the point-spread-function (PSF) applicable at the matching location on the ACIS detectors.   For consistency in the figure, we have normalized the post-shock peak (determined after smoothing with a Gaussian of FWHM$\,\approx \,$PSF) to a common level, and have shifted each profile in the horizontal direction so that the rise to 50\% of the peak occurs exactly at pixel 100.

To give a quantitative measure for a putative halo, we have measured the surface brightness averaged over three narrow angular ranges ahead of the shock: 0\arcsec--\,5\arcsec, 5\arcsec --\,10\arcsec, and 10\arcsec --\,15\arcsec.  For the zero point, we have taken the half-height point of the measured profile and have added the distance over which the appropriate PSF drops from its peak to 10\% of  its peak value.  For the local background, we have measured the average level starting 20\arcsec\ upstream from the shock and extending to 50\arcsec.   These measurements were carried out on the  exposure-corrected, flux-calibrated images (units photons cm$^{-2}$ s$^{-1}$ pixel$^{-1}$);  to obtain the uncertainties, we extracted profiles from the identical regions in the raw counts
images and obtained the relative uncertainties directly from Poisson
statistics.  The results are given in Table~4, and are also indicated in Figure~\ref{fig:halo_profiles}.  In the table, the post-shock peak value is that of 
the immediate post-shock peak (again after smoothing using a Gaussian with FWHM$\,\approx \,$PSF), 
and the peak/halo value is the ratio of this value to that of the net halo averaged over 0\arcsec--\,5\arcsec.

In five of the six cases we find a small excess over the local background for the X-ray flux 0\arcsec--\,5\arcsec\ ahead of the shock, though the statistical significance is low in all but two of the cases.  From that level, the X-ray flux jumps by well over an order of magnitude, over a distance comparable to the PSF width of 2\arcsec--\,4\arcsec, to the immediate post-shock peak.    Further upstream from the shock front, in the range 5\arcsec--\,10\arcsec, the excess over background has dropped to below $2\sigma$ (except for region E-2, the one which showed {\em no} excess over 0\arcsec--\,5\arcsec), and beyond 10\arcsec\ the flux is essentially indistinguishable from the background level in all six regions.   The region that shows the greatest evidence for a halo is E-1, which is  similar to that used by \citet{morlino10}, and which is also the one where the observation was made farthest off-axis.    Here the flux jumps across the shock by a factor $\sim 20$, and the preshock emission appears to follow near-exponential decay ahead of the shock, falling off by $1/e$ in about $4\arcsec \approx 1.3 \times 10^{17}$ cm .   
In the other cases, the decay scale is shorter, and/or the jump from halo to post-shock peak is greater.
We point out that there are a number of
factors other than a true halo that could give faint emission ahead of
the peak: curvature across the region, projection effects along
the line of sight, faint point sources that were not excised, and/or
intrinsic PSF response; yet there is no plausible way to make the
shock jump appear {\em sharper} than it really is.   
While we can by no means rule out the existence of a precursor halo, a fair summary of our results is that halo emission in the 1-4 keV range is typically narrower than 3\arcsec, and that across the shock the emission typically jumps by at least a factor of 20.

\subsection{Shock Models: Unmodified and Modified by Cosmic Rays}
 
For an unmodified shock, in which the pressure of upstream accelerated
particles is negligible, we expect a sudden density jump by a factor
of the compression ratio (4 for a nonrelativistic monatomic gas) at
the shock over a distance of a few thermal proton gyroradii, or about
$5 \times 10^9\;(d/2.2\,{\rm kpc})(v_{\rm shock}/5000\kms) (B/10 \,\mu{\rm
G})^{-1}\;$cm for SN~1006.   Upstream,
the relativistic particles will diffuse ahead of the shock a distance
of order $\kappa/v_{\rm shock}$, where $\kappa$ is the diffusion
coefficient (which may depend on particle energy as well as
orientation with respect to the magnetic field).  The flow velocity in
the shock frame is constant at the upstream value until suddenly
dropping due to viscous dissipation in the shock layer.  This viscous
dissipation heats the thermal gas and produces the obvious remnant
edge.   A precursor halo of X-ray emission would result from the upstream relativistic
particles radiating in the presumably constant magnetic field.

But for the case of a cosmic-ray modified shock, where efficient shock acceleration produces a non-negligible pressure from fast particles, the gross dynamics are changed: the inflowing gas (in
the shock frame) gradually slows over the diffusive scale length,
$\kappa/v_{\rm shock}$.  As a result, the overall gas compression  rises
gradually over that length scale, rather than abruptly, although a final
sharp jump at a remaining ``viscous subshock'' is expected.\footnote{While
some theoretical models predict that the viscous subshock will vanish altogether when particle acceleration is highly efficient, its presence in SN~1006 and other young remnants is clear from the
existence of hot thermal plasma.  Furthermore,  typical calculations, such
as \citet{ellison96} tend to find a minimum subshock compression ratio
of about 2.5 as part of an overall larger compression ratio that can
be quite large.}
For a modified shock, the region in which the
gradual deceleration and compression takes place is the shock
precursor. 
In this region, tangential
components of magnetic field would  grow simply due to compression.  In addition, the field can be amplified by large factors due to nonlinear effects \citep[e.g.,][]{bell04}.  
All these factors contribute
to the prediction of an X-ray synchrotron ``halo'' ahead of the
viscous subshock \citep{reynolds96}.   Figure~\ref{fig:halo_cartoon} illustrates these basic components in schematic form.

The expected scale length for any preshock emission would be somewhere
between the (unrealistic) minimum of the electron gyroradius $r_g$ for an unmodified, perpendicular
shock (without cross-field diffusion), and the diffusive 
scale length $\kappa/v_{\rm shock}$ for a parallel shock.  (Here
``perpendicular'' and ``parallel'' refer to the angle between the mean
upstream magnetic field and the shock normal.)  For a relativistic
electron, $r_g = eE/B$, and such an electron  emits the peak of
its synchrotron radiation at an X-ray energy $h\nu_0 = 7.5 E^2
B$ keV, giving $r_g = 2.4 \times 10^{16} (h\nu_0/{\rm
keV})^{1/2} (B/10\ \mu{\rm G})^{-3/2}$ cm.  For particles radiating
their peak at 4 keV in a magnetic field of 100 $\mu$G
\citep[somewhat higher than estimates for SN~1006 based on rim widths,
  e.g.,][]{parizot06}, $r_g = 1.5 \times 10^{15}$ cm, or about
$0.05''$ at 2.2 kpc---far too small to be detectable even from {\em
Chandra}.  

The diffusive  scale length is considerably longer, however.
For Bohm diffusion, the mean free path $\lambda$ is just $r_g$, and
$\kappa = \lambda v/3 = r_g c/3$, so the length scale is longer than the gyroradius by
$c/3v_{\rm shock}$.  For  5000 km s$^{-1}$ shock in SN~1006, the diffusive scale
is thus $20\,r_g$ or about $3 \times 10^{16}\; {\rm cm} \approx 1'' \approx
2$ ACIS pixels.  This is the solution
favored by \citet{morlino10}.  
Our data do not  rule out a halo this narrow.  
A halo on the $10''$ scale, well above what our data allow, would require (for the same factor of $c/3v_{\rm
shock} = 20$) $r_g \sim 1.5 \times 10^{16}$ cm for 4 keV electrons,
and hence $B \sim 22\,\mu$G.  Similarly, a halo on a $5''$ scale would
require $B \sim 34\, \mu$G.  Detection of a halo on these scales would
clearly demonstrate the presence of substantial magnetic-field
amplification in the shock precursor.   Profile E-2 may pose the most stringent constraints, as it appears to show no evidence of halo emission at all, within the limits of our observations.  A precursor narrower than 1\arcsec\ requires a precursor magnetic field of at least 79 $\mu$G.

Thus, synchrotron emission at 4 keV produced by electrons accelerated
in a cosmic-ray-modified shock with strong nonlinear magnetic-field
amplification \citep[e.g.,][]{bell04} can barely be accommodated within our
observations.  However, since the diffusion coefficient, and hence the
halo width, is expected to rise with energy, the halo could
become extended enough to observe for photon energies above 4 keV.
For long enough integration times, this might be possible with {\em
Chandra.}

If the halo scale is larger than about 1\arcsec, a large jump in
emissivity at the viscous subshock would be necessary to explain our
observations.  The particle distribution is continuous at the subshock
in virtually all models, so any emissivity jump would reflect the
magnetic field only.  For a parallel, modified shock, magnetic-field
amplification might take place gradually in the precursor, in which
case no jump in emissivity would be expected at the viscous subshock.
In this case, only the ``too narrow'' option is available to
accommodate our observations.  The synchrotron emissivity is
$j_\nu \propto B^{1 + \alpha}$ where $\alpha$ is the radio spectral
index.  For SN 1006, $\alpha \cong 0.55$ \citep{green09}, so the
maximum jump in $j_\nu$ occurs for a perpendicular shock, in which
case $B$ would rise by a factor equal to the compression ratio, in the
absence of additional downstream amplification processes.  For an
unmodified shock that factor is $4^{1.55} = 8.6$, not nearly enough to
explain the jumps we see. However, it is also possible that any
magnetic-field amplification would take place behind the
subshock \citep[e.g.,][]{giacalone07}, in which case an emissivity
jump at the subshock could be large.   If instead the shock is a modified one, the
overall compression ratio is larger, but the shock transition is now
broader, with the viscous subshock (with compression ratio $< 4$)
presumably accounting for the sudden emissivity jump---so the
predicted step at the subshock would be even smaller, and inconsistent
with our results.  Independent of models, a jump by a factor of $\gtrsim 20$ in emissivity
requires a jump in magnetic field by a factor of $\gtrsim 7$.

To summarize, we see minimal indication of emission beyond a sudden
steep rise which we presume to indicate the viscous shock.  Any
upstream emission is either confined to within $\lesssim 3\arcsec$ of that
rise, or fainter than about 1\% of the peak.  The ``too narrow" option
requires either a well-ordered magnetic field perpendicular to the
shock velocity at all of our profile locations \citep[disfavored by radio
data,][]{reynoso13}, or a magnetic field that is substantially amplified
in the precursor over expected ISM values.  The ``too faint" option requires
some process greatly increasing the magnetic field at the viscous
subshock, but not before.  An unresolved precursor, with the magnetic
field growing by (at least)  a factor of 7 over the far-upstream value, is the
most straightforward interpretation of our data.
 
\section{Discussion\label{sec:discussion}}

Previous work \citep[e.g.,][]{koyama95,dyer01, long03, yamaguchi08} has
demonstrated conclusively that almost all the thermal X-ray emission
in SN~1006 is due to ejecta, giving us the opportunity to learn about
the thermonuclear event that produced the remnant, and about
evolutionary processes operating early in the interaction of the
ejecta with the surrounding medium.  The datasets we present in this
paper allow us to locate the forward shock conclusively around most of the rim through the  \ha\ emission,
and to locate ejecta through the
 bulk of the thermal X-ray emission (though some O and Mg
may also arise from an ISM component).  A low density or low ionized
fraction can preclude Balmer emission, but where such emission is seen its presence indicates
the outermost shock, whether at the limb or in projection against the
interior.

\subsection{Outer Blast Wave and ISM Interaction}

The presence of Balmer filaments, though faint, around virtually
the entire limb indicates that there is at least partially neutral
material all around the periphery of SN 1006, and features projected
against the remnant allow us to identify material on the front or back
surfaces of the remnant in contrast to material in the interior.  We
locate the shock in the SE, where ejecta plumes appear to reach to
within 3\arcsec\ to  30\arcsec\ of the shock. We also identify bowshocks in
\ha\ and X-rays, the former evidently due to ejecta knots beyond the
mean shock surface, as seen extending beyond the remnant edge in several
locations and also in some interior locations (see Section 4 and Figure
\ref{fig:bowshocks}).  Cassam-Chena\"\i\ et al.~(2008) also remarked
on the bowshocks protruding beyond the mean blastwave location.
However, our deeper \ha\ image shows faint \ha\ emission at slightly
larger radii than was evident in the image they displayed
\citep[from][]{winkler03}.

We can take advantage of {\it Chandra's} superior spatial resolution to directly measure the thermal emission from the shocked ISM in the SE\@.  Both \citet{acero07} and \citet{miceli12} have reported an indirect spectroscopic detection of a shocked ISM component in addition to a shocked ejecta component, based on {\it XMM-Newton} spectroscopy.   But with {\it Chandra}, we can directly  separate out the two components {\em spatially}. We show in Figure~\ref{fig:south_protrusions} a region in the SE that was constructed to be outside of the fluffy ejecta structure, yet {\em inside} the extent of the faintest H$\alpha$ shock as seen in the optical image. This region, covering $\sim 30$ degrees of arc in length, ranges in thickness from 3\arcsec\ to 30\arcsec, with an average thickness of $\sim 20\arcsec$.

The background-subtracted spectrum from this region shows extremely  faint thermal emission, which we fit with an absorbed plane-shock model. For our purposes here, we are most interested in the emission measure ($\equiv n_{e}n_{p}V$, where $n_{e}$ and $n_{p}$ are the electron and proton densities, respectively, and $V$ is the volume of the emitting region), which we find to be 3.7 $\times 10^{54}$ cm$^{-3}$. We estimate the line-of-sight depth through the emitting region to be half the length of the region, making $V \approx 9 \times 10^{55}\;{\rm cm}^3$. Assuming cosmic abundances, where $n_{e} = 1.2\,n_{p}$, and a filling fraction for the gas of unity, we obtain a mean post-shock proton density of $0.18 ^{+0.20}_{-0.08}\: {\rm cm}^{-3}$, temperature $kT \approx 0.80$ keV, and ionization timescale $n_e\,t \approx 3.9 \times 10 ^8\: {\rm cm^{-3}\,s}$. Assuming the standard compression ratio for a strong shock of 4, this leads to a pre-shock density of $n_{0} = 0.045 ^{+0.049}_{-0.020}\: {\rm cm}^{-3}$, comparable with the results obtained in the {\it XMM-Newton} analyses.  Cosmic-ray modification of the shock, which \citet{miceli12}  found to be consistent with their spectral fits along the SE rim, would lower this pre-shock density, since it would raise the compression ratio of the shock.

While the value obtained above  for  the ISM density along the SE rim is low, it is not surprisingly so for a region far above the Galactic plane, and is consistent with previous preshock density determinations for SN~1006 \citep[for regions other than along the  NW bright  filament, where considerably higher preshock densities have been measured, $n_0 \approx 0.15-0.3\; {\rm cm}^{-3}$,][and numerous references therein]{winkler13}.   \citet{long03}  estimated an upper limit of
0.1 cm$^{-3}$ in the NE synchrotron-dominated limb, consistent with
the value we find here for the SE\@.
\citet{miceli12} fit data from deep  {\em XMM-Newton} observations of the SE region using a multi-component shock model with two non-equilibrium ionization components (model {\tt vpshock} in XSPEC) plus a non-thermal component, to spectroscopically separate the ejecta (with variable abundances) from the shocked ISM (abundances fixed at solar).   From the solar-abundance component, they estimated a preshock density of $\lesssim 0.05\:{\rm cm}^{-3}$, consistent with our results and with an earlier {\em XMM-Newton}  analysis by \citet{acero07}.   
In view of the potential ambiguities that attend analyses using multiple components (including those by Long \etal, Acero \etal, and Miceli \etal), our analysis of
emission that originates beyond the ejecta plumes should provide a cleaner
limit on the ISM density.


\subsection{Large-scale Ejecta Distribution}

The images of Figure~\ref{fig:eqw} show immediately that the ejecta
distribution is not symmetric, with Si much brighter in the SE than in
the NW, confirming the {\em Suzaku} result recently reported by
\citet{uchida13}.  As in that paper, we see that O is distributed more
uniformly than Si.  Indeed, for the bright X-ray filament along the
NW, we found in \citet[][based on a reanalysis of the archival data used by Long et al.\ 2003]{winkler13} that solar abundances can describe
the X-ray emission, indicating a significant contribution from interstellar
oxygen.  However, Si is supersolar over most of the remnant, as
expected for SN Ia ejecta; its highly asymmetric distribution
indicates substantial asymmetries in the ejecta in general.  While
ionization effects can also cause variations in the strength of Si
lines, suppressing Si K$\alpha$ emission in the NW as much as we
observe would require a far lower ionization timescale than observed
there \citep[$\sim 2 \times 10^9$ cm$^{-3}\,$s,][]{long03}.
Nonuniform distributions of emission from O, Ne, and Mg,  as
well as reported distributions of Fe from {\em Suzaku}
\citep{yamaguchi08}, all support the necessity for asymmetric ejecta.

Asymmetries, attributed to the explosion itself,  are also indicated from UV absorption-line spectroscopy of UV-bright point sources behind SN1006 \citep[e.g.,][]{hamilton97, winkler05, hamilton07, winkler11}. These spectra show large differences in the column density of \feii\ in the freely expanding  ejecta along lines of sight separated by several arcminutes, as well as strongly asymmetric profiles of \feii\ and \siii\@.  In particular, a sharp red edge in the \siii\,$\lambda$1260 absorption profile for the \citet{schweizer80} star indicates that the fastest unshocked ejecta (i.e., material  just encountering the reverse shock) on the far side of the SNR is traveling outward at $7000\kms$, but that such  ejecta on the near side is moving far slower.

We therefore consider models for the distribution of ejecta in SN~Ia explosions.  
Various authors  \citep[e.g.,][]{kasen09,
  maeda10, seitenzahl13} have demonstrated in recent years that 2-D
and 3-D calculations of Type Ia explosions can produce somewhat different results from
those of classic spherically symmetric models such as W7 of \citet{nomoto84}---both in nucleosynthetic yields and in kinematic
distribution.  The 3-D delayed-detonation models of
\citet{seitenzahl13} predict a range of O and Si ratios and locations.
After about 100 seconds, ejecta are in ballistic motion (i.e.,
pressure forces are negligible and material is freely expanding).  At
that time, for the most asymmetric models (ones with the fewest ignition points),
azimuthally-averaged O and intermediate-mass elements have
similar distributions, while for more symmetric models, O is found at
considerably larger distances than Si. The strong Si asymmetry we find
is most easily explained by their very asymmetric model N3 (see their
Figure~3); we would then interpret our more uniform O distribution as
an indication of substantial shocked-ISM oxygen.  Mg is intermediate;
the models of \citet{iwamoto99} produce 3 to 8 times as much Mg as Ne,
so we might expect to find a substantial component of Mg from ejecta in addition to some that has resulted from shocked ISM.
All this is consistent with the
visual impression from Figure~\ref{fig:eqw}.  We conclude that the
ejecta distribution in SN 1006 qualitatively supports quite asymmetric
SN Ia models, though detailed quantitative spectral analysis and
modeling will be required to make this statement more definite.

\subsection{Small-scale Ejecta Distribution}\label{disc:small}

The arcminute-scale ``puffy'' structure of ejecta apparent in SN 1006
X-ray images has been noted by many authors.  Its origin could be in
intrinsic clumpiness produced in the explosion itself (``intrinsic''
clumps) or in hydrodynamically produced structures from
Rayleigh-Taylor instabilities at the contact interface (R-T clumps).
Intrinsic clumps could arise if the ejecta are subject to the ``nickel
bubble'' effect \citep{li93} in which $\gamma$-rays from the radioactive
decay of $^{56}$Ni cause local expansion of the ejecta and sweep
surrounding ejecta into a shell, which is then fragmented by
Rayleigh-Taylor (R-T) instabilities in the first few minutes after the
explosion.  Later developing R-T clumps are seen in all hydrodynamic simulations
of SNR evolution \citep[e.g.,][]{chevalier92, jun96, orlando12, warren13}
Determining which
process produces the clumpy structure in SN 1006 is important, as
intrinsic clumps can be used as diagnostics of the explosion, while
R-T clumps contain information on the global hydrodynamic evolution.
Unfortunately, the two models typically produce structures of similar
appearance: R-T clumps are formed from dense ``mushroom caps'' of less
decelerated material, and thus appear disk-like, while intrinsic clumps are
flattened to a similar shape as they are slowed and eventually
fragmented by instabilities \citep{wang01, orlando12}. Such
morphologies would result in structures much more easily visible from
the side (due to longer lines of sight through material) than from face on,
and can explain the lack of small-scale \ha\ emission toward the
center of SN 1006.

Our observations of numerous bowshock structures in
\ha\ (Section~\ref{features}) illustrate that it is relatively common for ejecta clumps to reach
the outer blast wave and interact with at least partially neutral
material.  Such structures are often described as ejecta
``bullets'' or ``shrapnel'' \citep[e.g.,][]{wang01, miceli13a}.  Those
seen in projection against the remnant interior (see
Figure~\ref{fig:bowshocks}) present serious challenges to X-ray
spectral analysis due to projection effects.  Particularly useful are
the structures that can be seen at the extreme edge of the  remnant, where they
clearly extend beyond the mean blast-wave radius.  

The very existence
of such protrusions poses significant difficulties for models.  The
overall puffy structure of ejecta in SN 1006 strongly resembles the
simulated images from 3-D hydrodynamic simulations by
\citet{warren13}.  Those simulations evolved an initial exponential
density profile into a uniform ambient medium; the ejecta structure
was produced purely by hydrodynamic instabilities.  But the
simulations of \citet{warren13} are able to produce clumps breaking
through the mean shock radius only if the shock compression is quite
high (simulated by an artificially low adiabatic index, $\gamma = 6/5$
giving a compression ratio of 11), explained by efficient cosmic-ray
acceleration.  A high compression
ratio provides a less likely explanation in the SE, where the relativistic-particle
population is evidently less important, judging by the relative
weakness of nonthermal emission at radio, X-ray, and gamma-ray
wavelengths.

Intrinsic clumpiness may also be able to
produce similar projections, however.  If such clumps are sufficiently dense \citep[][find a density constrast of $\sim 100$ is
required for structures seen in Tycho's SNR]{wang01}, they would be less
decelerated and capable of producing the structures we see in SN~1006.
In the MHD simulations of \citet{orlando12}, magnetic-field amplification at the outer edges of clumps stabilizes these structures and enables some clumps with a density contrast of a factor of only $\lesssim 10$  to reach or surpass the mean blast-wave radius, without resorting to unusually compressive shocks.  The connection between the close approach of the contact discontinuity to the forward shock and efficient shock acceleration, as was favored by \cite{cassam-chenai08},  is no longer required.  Detailed spectral
analysis of features beyond the blast wave, and of sufficiently bright
X-ray structures behind bowshocks seen against the interior, may allow
a discrimination between the intrinsic vs later-time R-T models.

\section{Summary and Conclusions \label{sec:conclusions}}

Here we have provided an overview of new X-ray and optical observations of the remnant of SN~1006.  X-ray emission in SN~1006 is complex:  the interior of the SNR is filled with small-scale features arising primarily from SN ejecta, while the shell rims  exhibit both synchrotron radiation from electrons accelerated at the shock front (in the NE and SW), and thermal X-ray emission from hot plasma---both SN ejecta and shocked ISM---in the NW and SE.  Our primary results are as follows:

\begin{itemize}
\item
\HA\ emission can be traced around almost the entire SNR shell, even in regions that are dominated by synchrotron radiation.  Very faint, diffuse \HA\ emission, arising from the near and/or far side of the SNR,  covers a substantial portion of the interior of the the SNR.  Wherever it is found, the short lifetime for neutral H atoms behind fast shocks like those in SN~1006 requires that the Balmer-line emission must occur immediately behind a shock encountering partially neutral ISM.

\item
Some of the small-scale (10\arcsec - 30\arcsec\ = 0.1 - 0.3 pc) X-ray features within the (projected) SNR shell  have associated Balmer filaments that resemble bow shocks.  The X-ray spectra of these tuft-like features shows that they are ejecta and that they have ionization timescales that are short compared to the time since the SN exploded.  The fact that these features have associated Balmer emission indicates that they are ejecta ``bullets'' that are penetrating the interstellar shock and encountering pristine ISM.  Further analysis will be required to determine whether these ejecta knots are a result of density inhomogeneities originating in the SN explosion, or have been produced  later through Rayleigh-Taylor instabilities as the SNR has evolved.  

\item
The expansion velocity of the outer edge of the SNR, as measured from the proper motion, varies dramatically as a function of azimuth.  The lowest  velocity is about $3000 \kms$ in the NW, where the strongest \HA\ emission occurs, and where the pre-shock density is highest.  In the SE, the velocity is almost 2.5 times higher, $\sim 7400 \kms$.   The synchrotron-dominated limbs in the NE and SW both have velocities of about 5000 $\VEL$. 

\item
The overall distribution of ejecta material is  asymmetric.  We confirm the results of \citet{uchida13}  that Silicon, which arises almost entirely from SN ejecta, is strongly concentrated in the the SE quadrant.  Emission from Neon is by far the strongest along the NW rim, where the primary shock is encountering denser material than anywhere else around the periphery.  Oxygen and Magnesium show less extreme concentration in the SE than does Silicon, and also show concentrations well inside the shell rim in the NW quadrant; these probably arise from a mix of SN ejecta and shocked ISM.  

\item
Our data place significant constraints on a possible X-ray halo in front of any of the synchrotron-dominated regions along the NE or SW limbs. We observe abrupt jumps in emission by a factor ranging from $\sim 20$ to $> 100$  over scales comparable with the PSF for the instrument at multiple locations.  Immediately preceding these jumps, there is slight evidence for a faint precursor on scales of $\lesssim 3\arcsec$.  The most straightforward explanation of these results is that diffusive particle acceleration is promoted by a magnetic field that is amplified by a factor of 7 or more in a narrow precursor region.

\item
Even in the new, deep {\em Chandra} images, there is no clear evidence for the primary shock along the rim of the SNR shell in the SE.  Instead, the X-ray structure there consists of a series of tufts, whose kinematics suggest that they have been  decelerated little if at all (expansion index $m \approx 0.8$) and whose spectra show they are ejecta-dominated.   Within a narrow region ahead of the X-ray tufts, but behind the outermost \ha\ emission, we find extremely faint thermal X-ray emission whose emission measure suggests a pre-shock density $n_0 \approx 0.045\:{\rm cm^{-3}}$, similar to the value inferred by other investigators through different arguments.  
\end{itemize}
 
 The data set from the {\em Chandra} Large Project to survey SN~1006 is a rich one, and should provide a resource for many future studies, by ourselves and others.

\acknowledgements

We acknowledge several suggestions from the anonymous referee that have led to greater clarity in this paper.   Support for this work was provided by the National Aeronautics and Space Administration 
through \chandra\ Grant Number GO2-13066, issued by the \chandra\ X-ray Observatory Center, 
which is operated by the Smithsonian Astrophysical Observatory 
for and on behalf of NASA under contract 
NAS8-03060\@.   
PFW also  acknowledges financial support from the National Science Foundation 
through grant AST-0908566.

\bibliographystyle{apj}



\include{tab1}

\include{tab2}

\include{tab3}

\include{tab4}\label{tab:halo_stats}

\clearpage

\begin{figure}
\epsscale{0.80}
\plotone{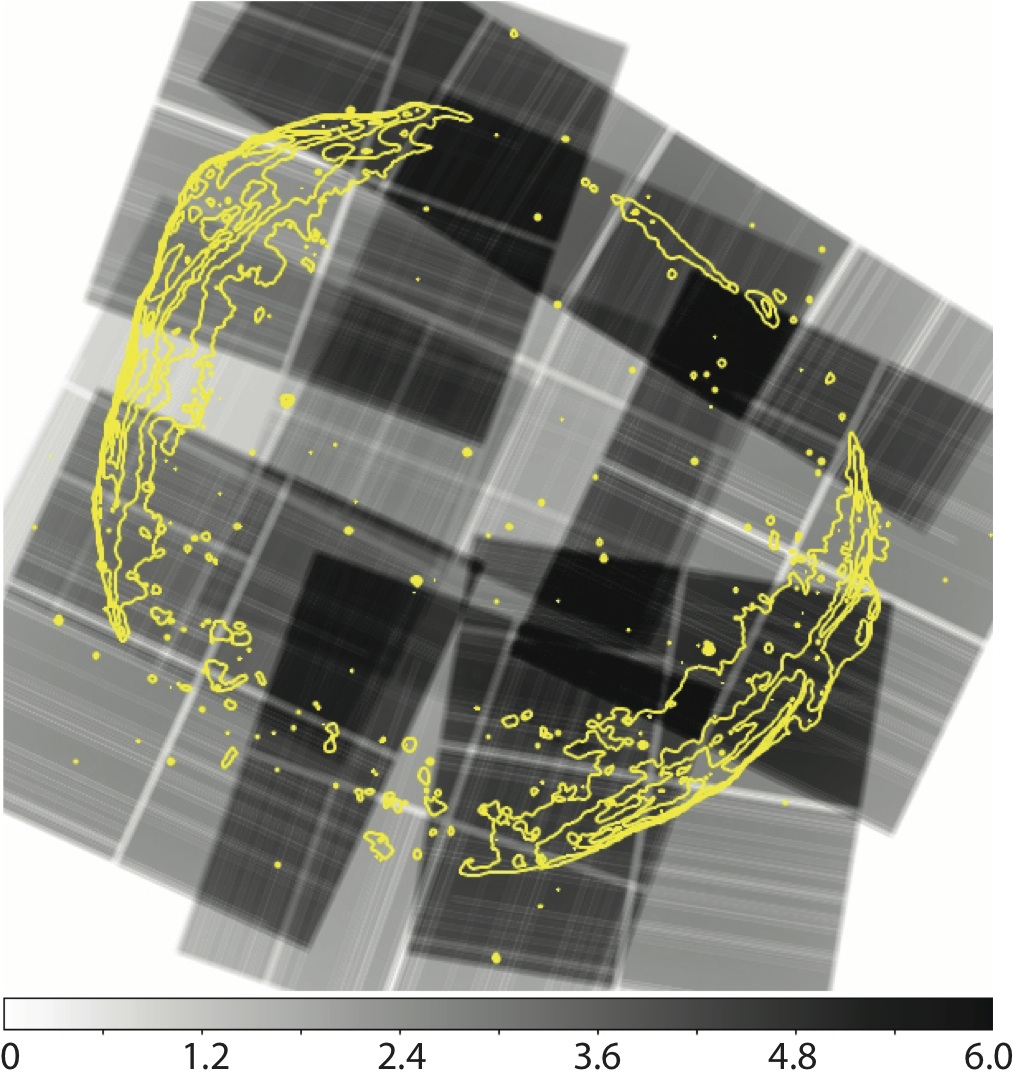}
\caption{Exposure map for the 2012 ACIS observations.  Contours representing the broad X-ray flux are overlaid.  The exposure time in most of the individual pointings ranges from 80 ks to 100 ks, but the exposure is, of course, deeper where multiple pointings overlap.  The scale gives the exposure in units $10^7 {\rm cm^2\;s}$.   
} 
\label{fig:xray_expmap}
\end{figure}

\begin{figure}
\epsscale{1.0}
\plotone{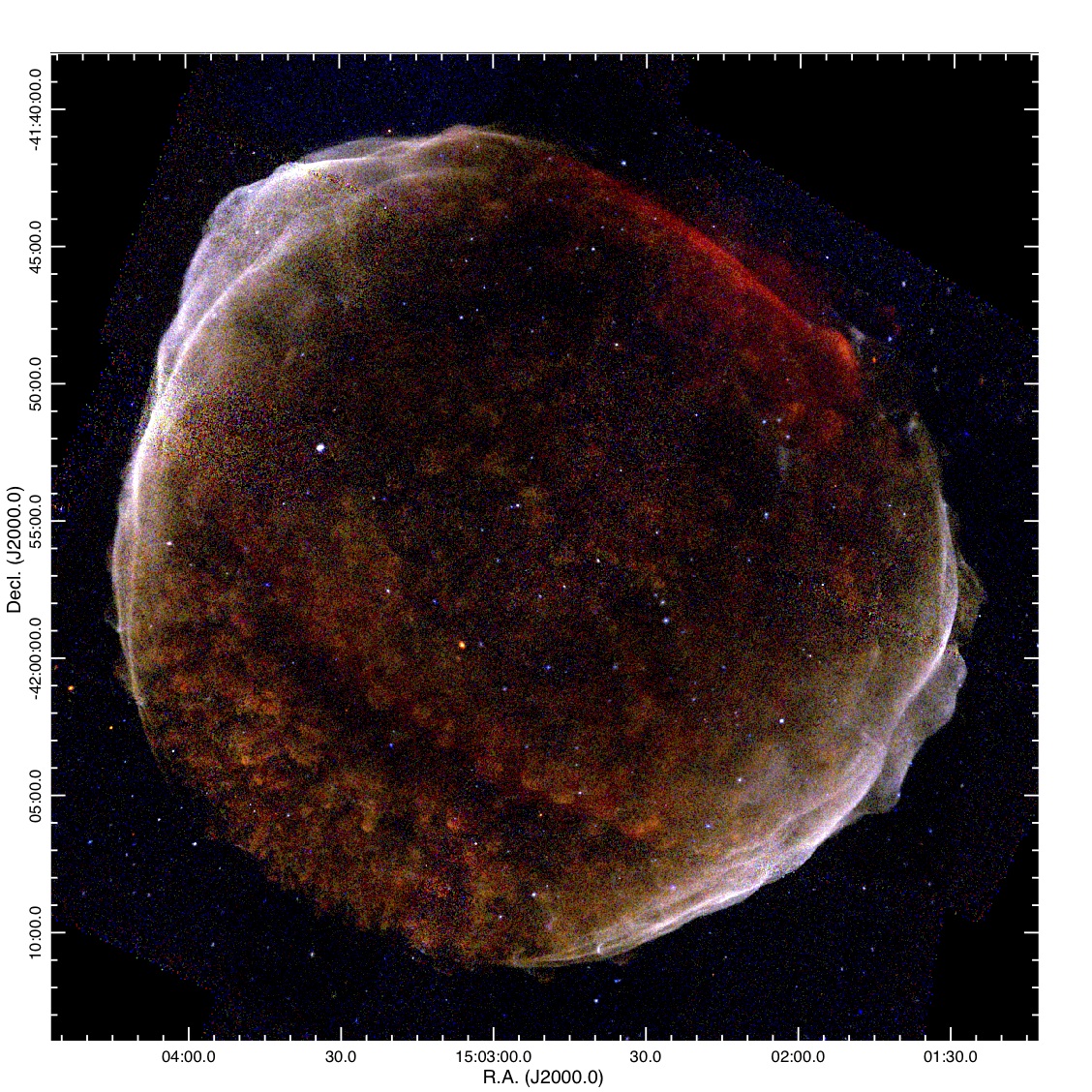}
\caption{``True-color" X-ray mosaic of all the 2012 \chandra\ ACIS observations of SN~1006; red = soft (0.5-1.2 keV), green = medium (1.2-2.0 keV), blue = hard (2.0-7.0 keV).  The synchrotron-dominated regions along the NE and SW rims are much harder than the thermal-dominated emission from elsewhere in SN~1006.  
} 
\label{fig:xray_rgb}
\end{figure}

\begin{figure}
\epsscale{0.8}
\plotone{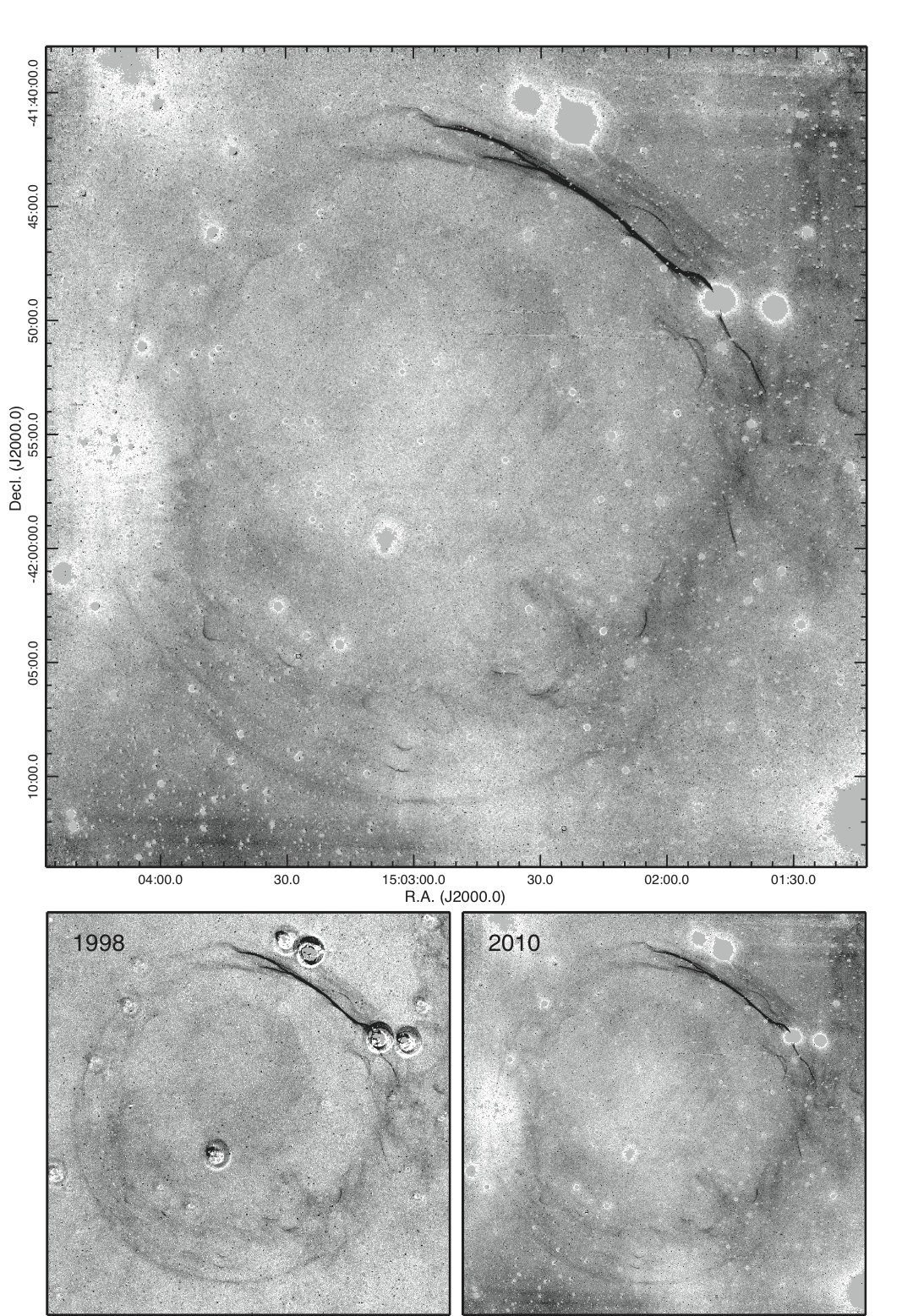}
\caption{The large upper panel shows a very deep \ha\ image of SN 1006, after continuum subtraction, obtained at the CTIO 4m Blanco telescope with the Mosaic II camera, 2010.  The relatively bright filaments to the NW are saturated in this display, in order to emphasize the far fainter emission elsewhere in the remnant.  The field is 36\arcmin\ square, and exactly matches that of the  X-ray image, Figure \ref{fig:xray_rgb}.   The smaller images below show both the  1998 and 2010 images  at the same scale; most of the features seen in the 2010 image are also visible in the earlier low-resolution image.
}
\label{fig:ha_image}
\end{figure}

\begin{figure}
\epsscale{0.70}
\plotone{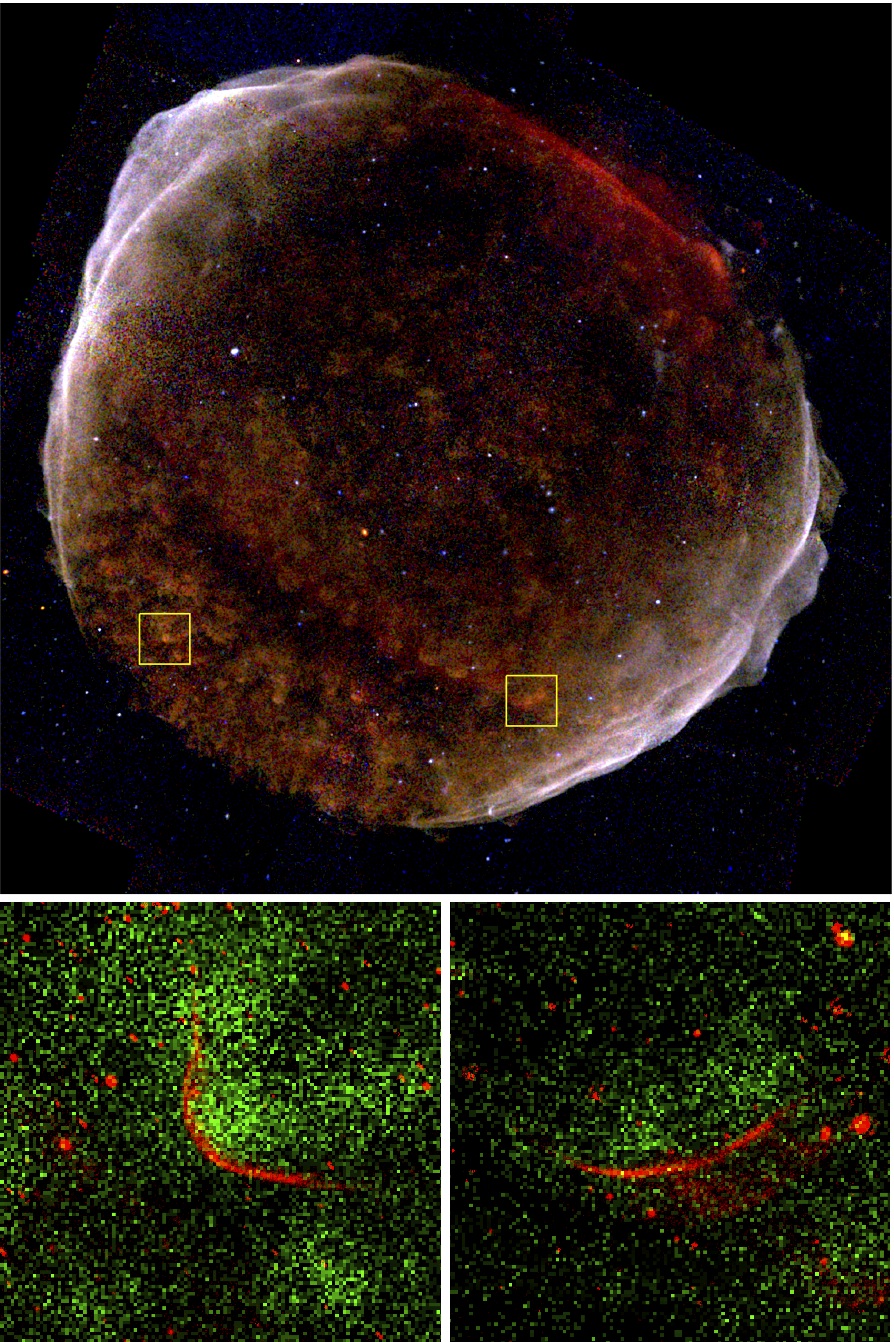}
\caption{Some of the bright tufts of soft X-ray emission appear immediately behind what appear to be bowshocks seen in \ha; the two lower panels show two examples, from locations indicated by the yellow boxes in the upper panel.  In the lower panels, \ha\ emission is shown in red, and 0.5-7.0 keV X-rays in green.  Both these panels are 2\arcmin\ square.   
} 
\label{fig:bowshocks}
\end{figure}

\begin{figure}
\epsscale{0.45}
\plotone{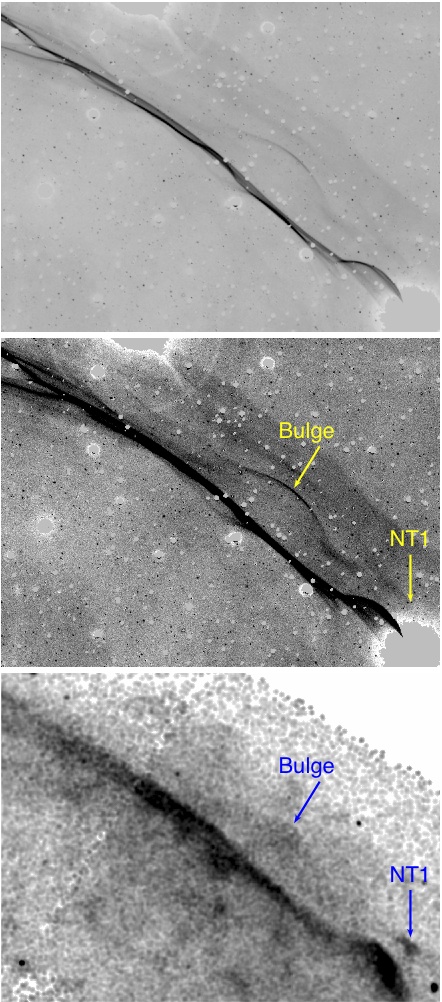}
\caption{The same $10\arcmin \times 7\farcm5$ region along the NW shock is shown these images:  the top and middle are in  \ha, displayed at different greyscale levels.  The upper image shows the delicate bright filament, while the middle one shows much fainter emission farther out, indicating the complex three-dimensional structure of the shock front.  The lower panel shows the soft (0.5-1.2 keV) X-ray image, displayed to show the coincident faint emission to the NW.   The bulge ahead of the bright \ha/X-ray filament is easily seen in both \ha\ and X-rays.  The feature marked NT1 is a nonthermal filament noted by \citet{katsuda13} whose proper motion indicates a velocity much higher than that for the brighter thermal filament in the NW; faint optical emission is also seen just ahead of this filament, though  this is partially lost to the bright halo around a bright star in this region.
} 
\label{fig:nw_shock}
\end{figure}

\begin{figure}
\epsscale{0.90}
\plotone{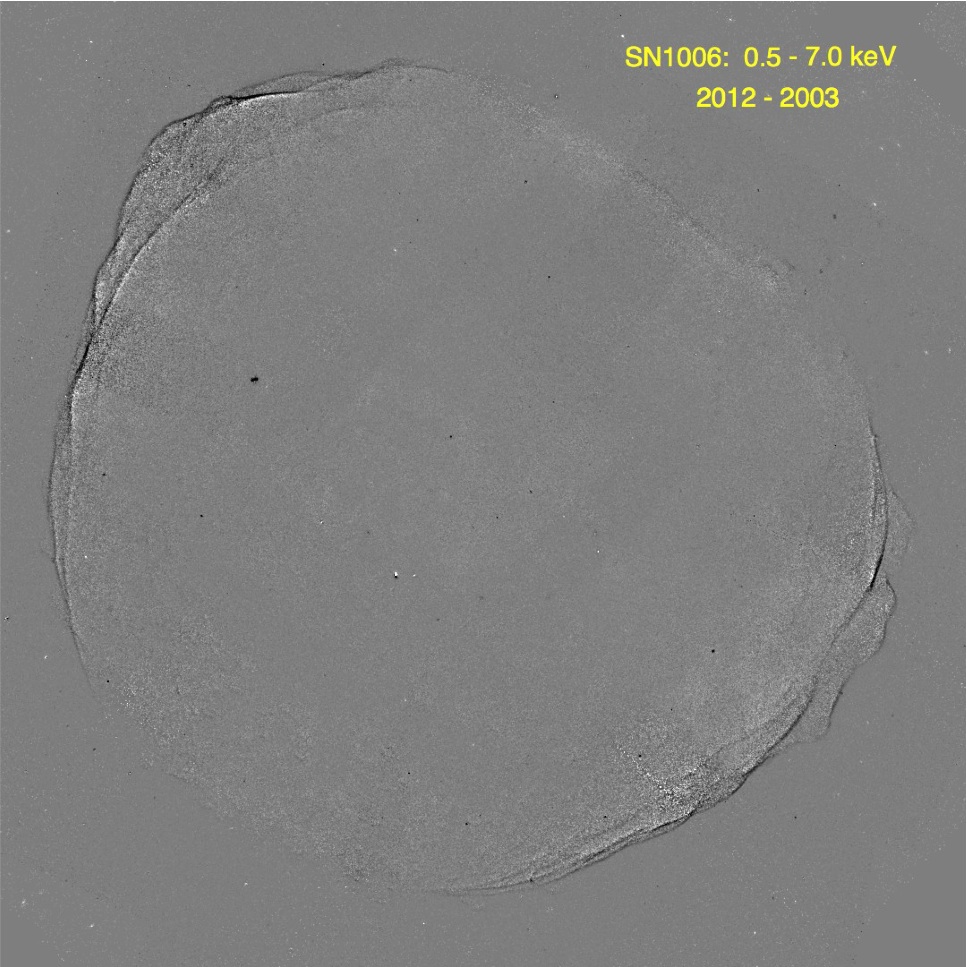}
\caption{Difference image between the merged 2012 image (0.5 - 7.0 keV, the sum of all three bands from Figure \ref{fig:xray_rgb}, shown as black here) and the 2003 one \citep[][shown as white]{cassam-chenai08}.  The 2003 data were aspect-corrected to match those from 2012, and then merged through a process identical to that for 2012.  Expansion is especially evident along the sharp NE and SW limbs, but is noticeable around almost the entire perimeter.  The point source that located just SE of the geometric center, which appears to have moved slightly southward, corresponds to a foreground star with high proper motion, and was not  used in the image registration.   An animated version of this figure, which also includes an X-ray/optical comparison, will appear in the on-line edition.
} 
\label{fig:xray_diff}
\end{figure}

\begin{figure}
\epsscale{0.9}
\plotone{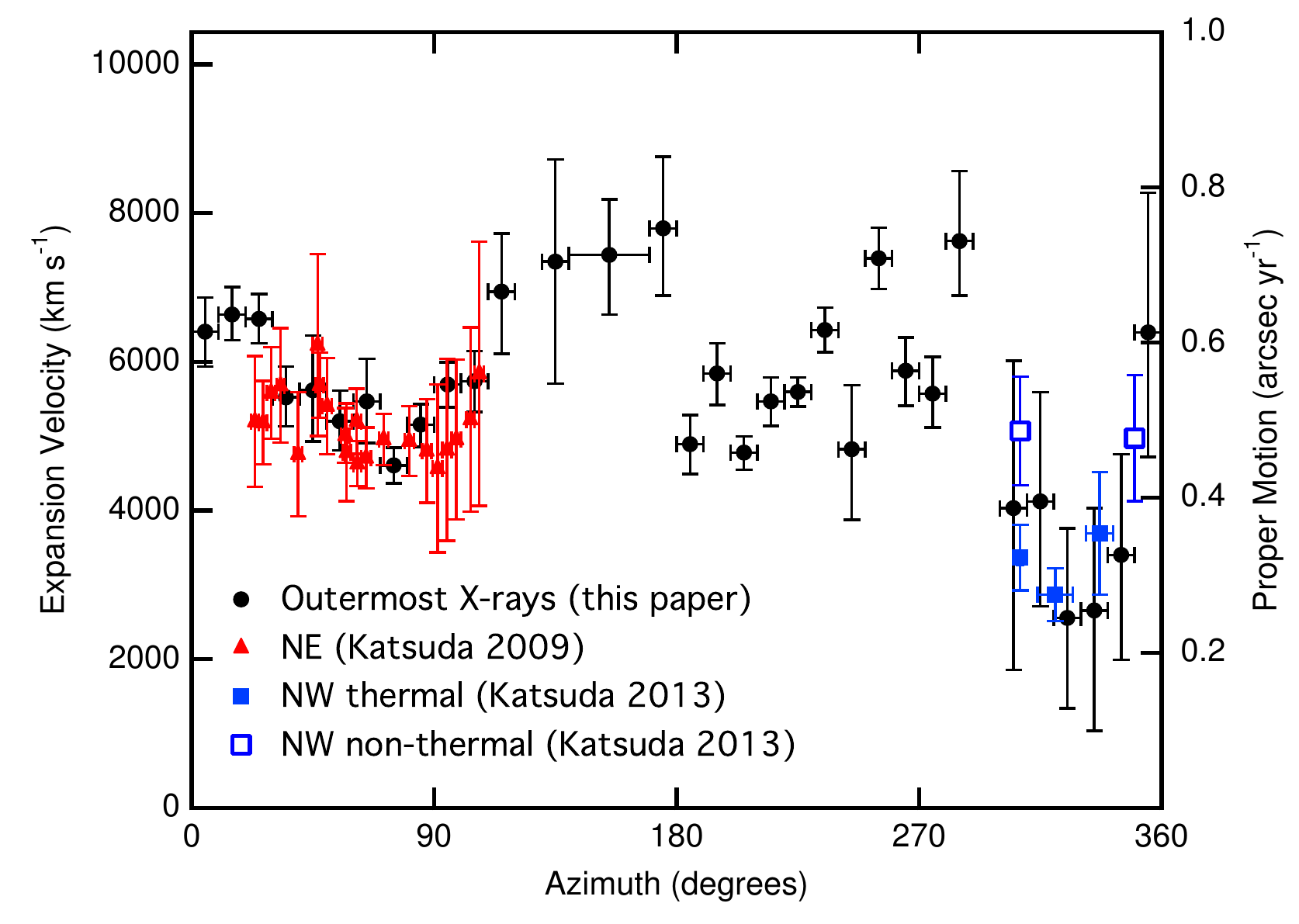}
\caption{The black circular points show the measured expansion of the SN~1006 X-ray limb from 2003 to 2012,  plotted as a function of azimuth (defined as counterclockwise from north).   Also shown are the measurements based on individual features in the NE \citep[red triangles,][]{katsuda09} and the NW \citep[in blue,][]{katsuda13}, where there are a few small non-thermal features (open squares) in addition to the thermal ones (filled squares).  For the expansion velocity, we assume a distance of 2.2 kpc.  All the uncertainties indicate 90\% confidence limits.
} 
\label{fig:azimuth_plot}
\end{figure}

\begin{figure}
\epsscale{0.9}
\plotone{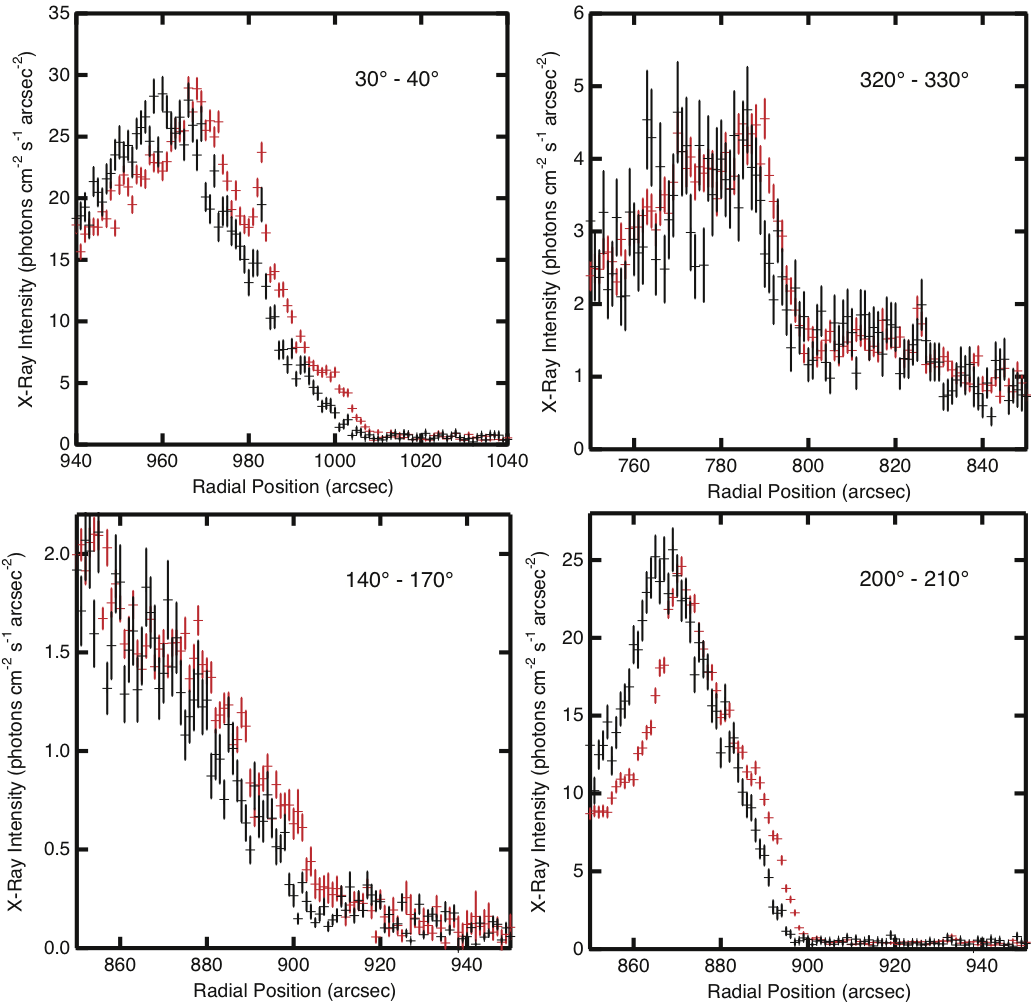}
\caption{Radial profiles for the 0.5-8 keV X-ray emission,  extracted from sectors of the SN~1006 shell at the indicated azimuth ranges, one taken from each quadrant.  Data points in black and red represent the 2003 and 2012 epochs, respectively.  In each case the proper-motion was measured by fitting the radial range including only the outermost X-ray emission.
} 
\label{fig:pm_profiles}
\end{figure}

\begin{figure}
\epsscale{0.9}
\plotone{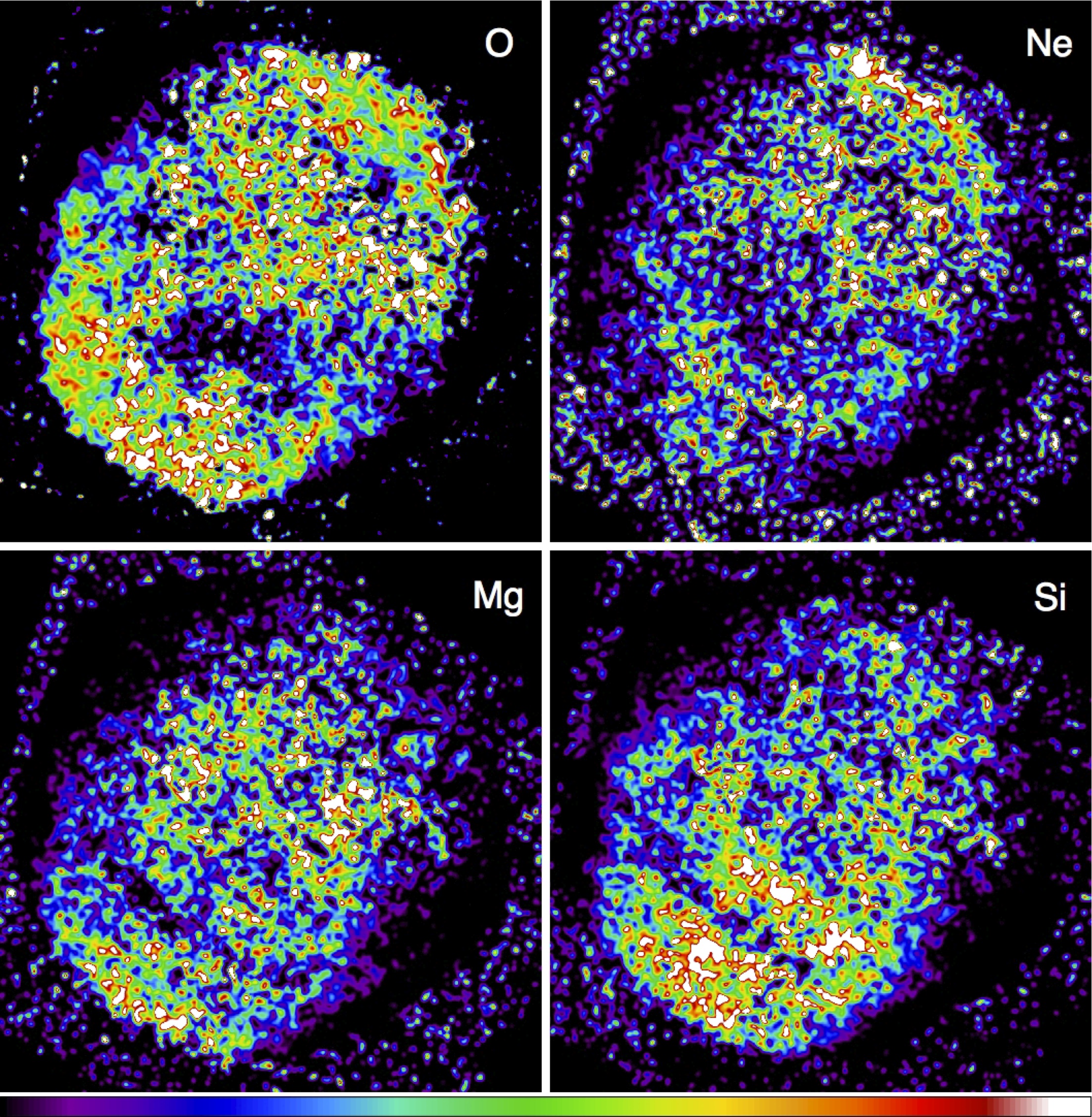}
\caption{Equivalent-width images in the K-lines of O, Ne, Mg, and Si, produced according to the procedure described in the text.  The intensity scales are linear, starting from 0 in each case with a maximum of 1 keV (O), 0.12 keV (Ne); 0.25 keV (Mg); 0.6 keV (Si).  The field is 36\arcmin\ square, oriented N up, E left.   An excess of line emission in the SE quadrant is evident in Si especially---an indication of an asymmetry in the ejecta distribution.  Virtually no line emission is seen from the crescent-shaped regions on the NE and SW limbs, since these regions are completely dominated by nonthermal emission.
} 
\label{fig:eqw}
\end{figure}

\begin{figure}
\epsscale{1.0}
\plotone{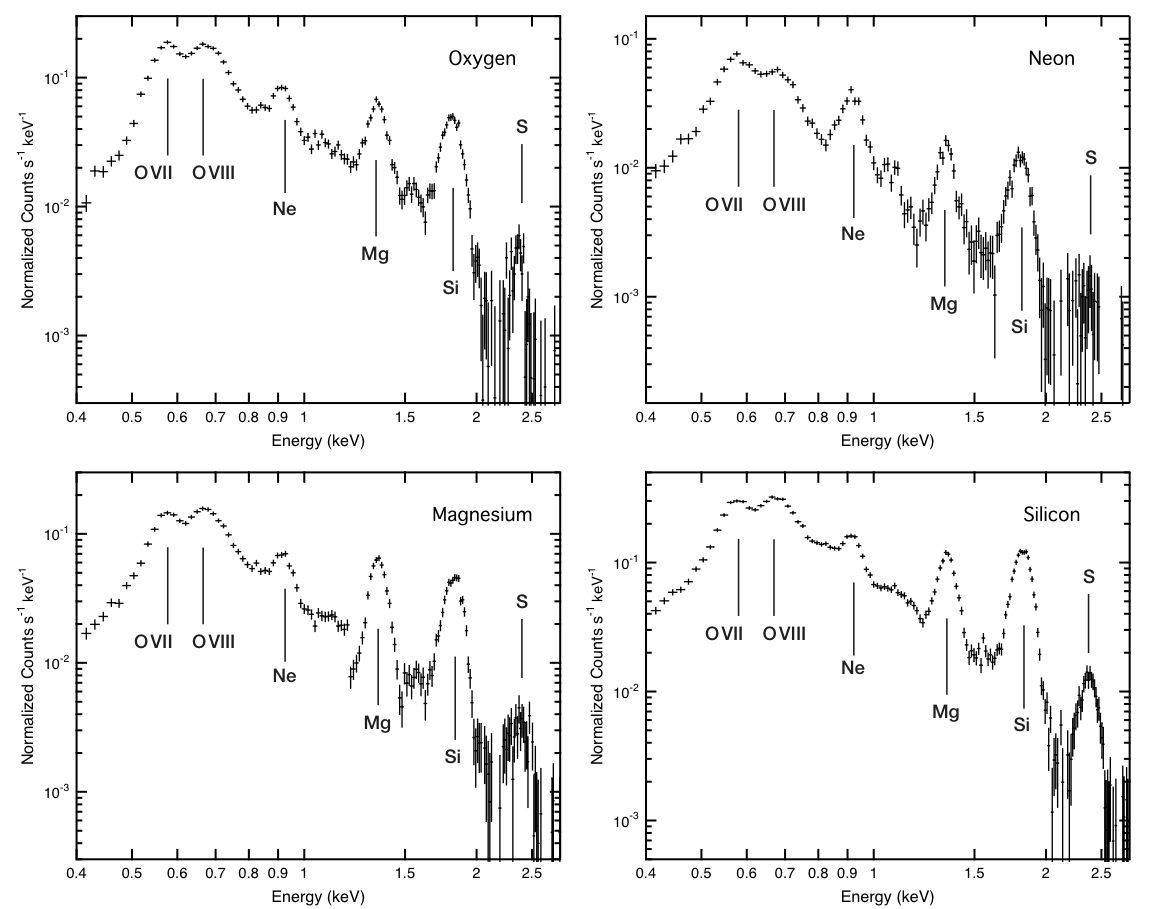}
\caption{X-ray spectra extracted from the brightest 25\% of the pixels in each of the equivalent-width images (Figure \ref{fig:eqw}).  
} 
\label{fig:spectra}
\end{figure}

\begin{figure}
\epsscale{0.8}
\plotone{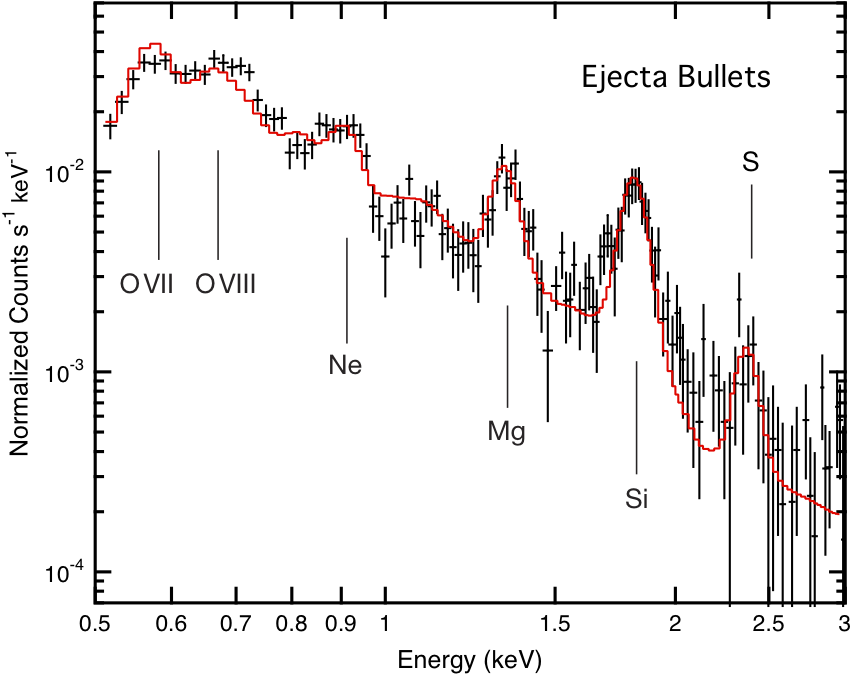}
\caption{Combined spectrum of the two ejecta bullets shown in Figure~\ref{fig:bowshocks}, with local background subtracted.  The best fit with an absorbed VNEI model is shown in red.} 
\label{fig:bullets}
\end{figure}

\begin{figure}
\epsscale{0.8}
\plotone{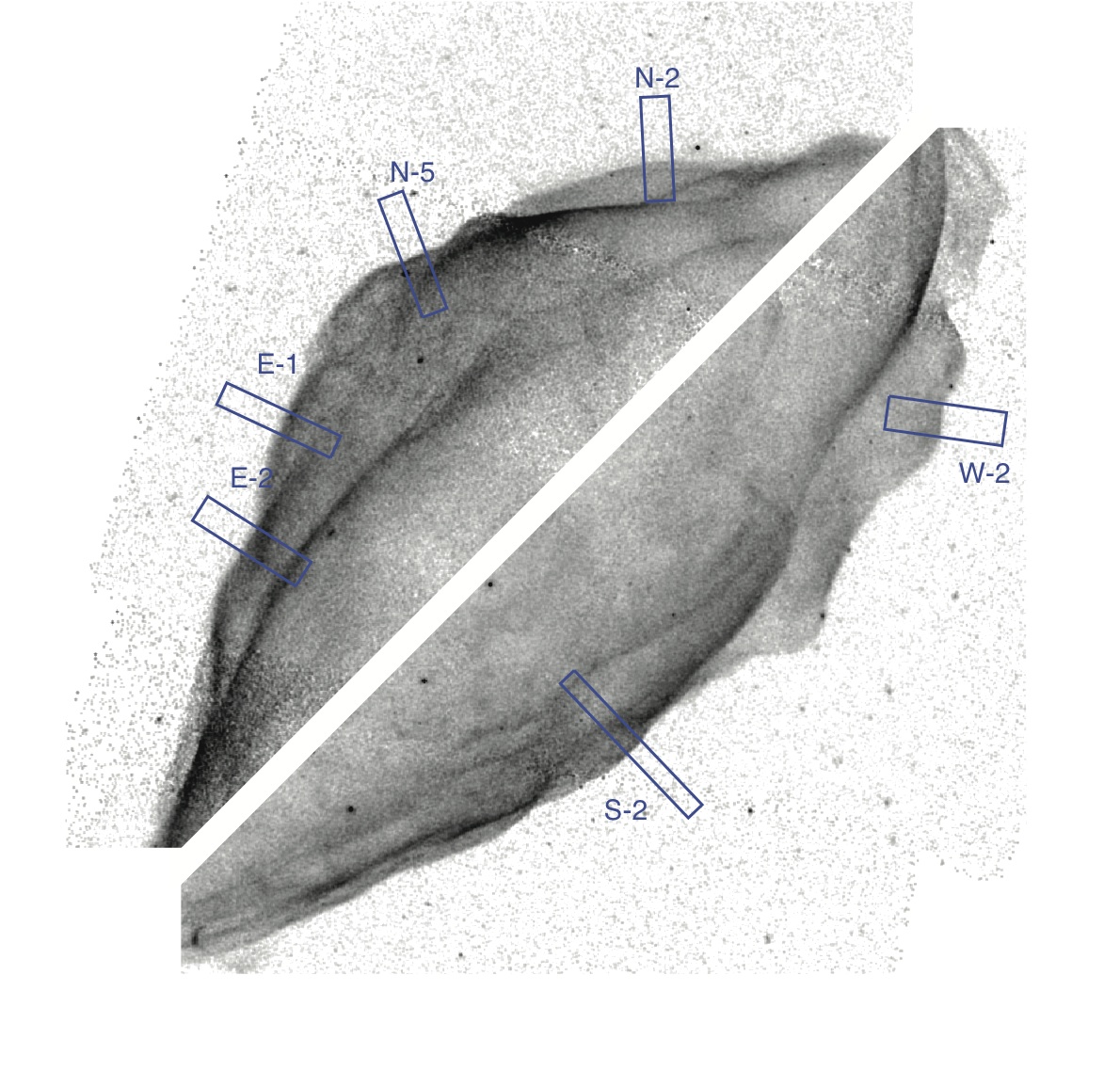}
\caption{Composite showing the NE (upper left) and SW (lower right) synchrotron-dominated limbs in the 0.5-7 keV mosaic image of SN\ 1006.  The blue rectangles indicate regions where the precursor profiles plotted in Fig.~\ref{fig:halo_profiles} were taken.} 
\label{fig:halo_regions}
\end{figure}

\begin{figure}
\epsscale{0.9}
\plotone{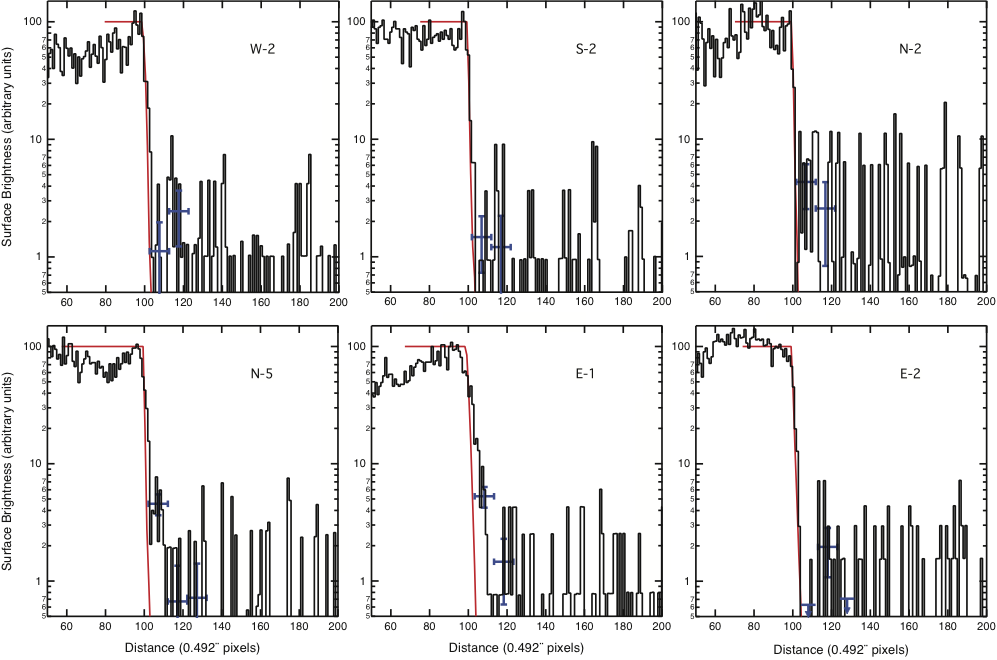}
\caption{Profiles across the shock front drawn from the six regions shown in Fig.~\ref{fig:halo_regions}---upstream to the right, downstream to the left, after subtracting the far-upstream background.  In each case, the profile has been normalized so that the immediate post-shock peak has been normalized to 100, and a horizontal shift applied so that the jump to half the post-shock peak occurs at pixel 100.   The red curves show the PSF response to a sharp edge at the shock position, and the blue points with error bars show the average X-ray halo level in different 5\arcsec\ intervals ahead of the shock (data from Table~4).
} 
\label{fig:halo_profiles}
\end{figure}

\begin{figure}
\epsscale{0.6}
\plotone{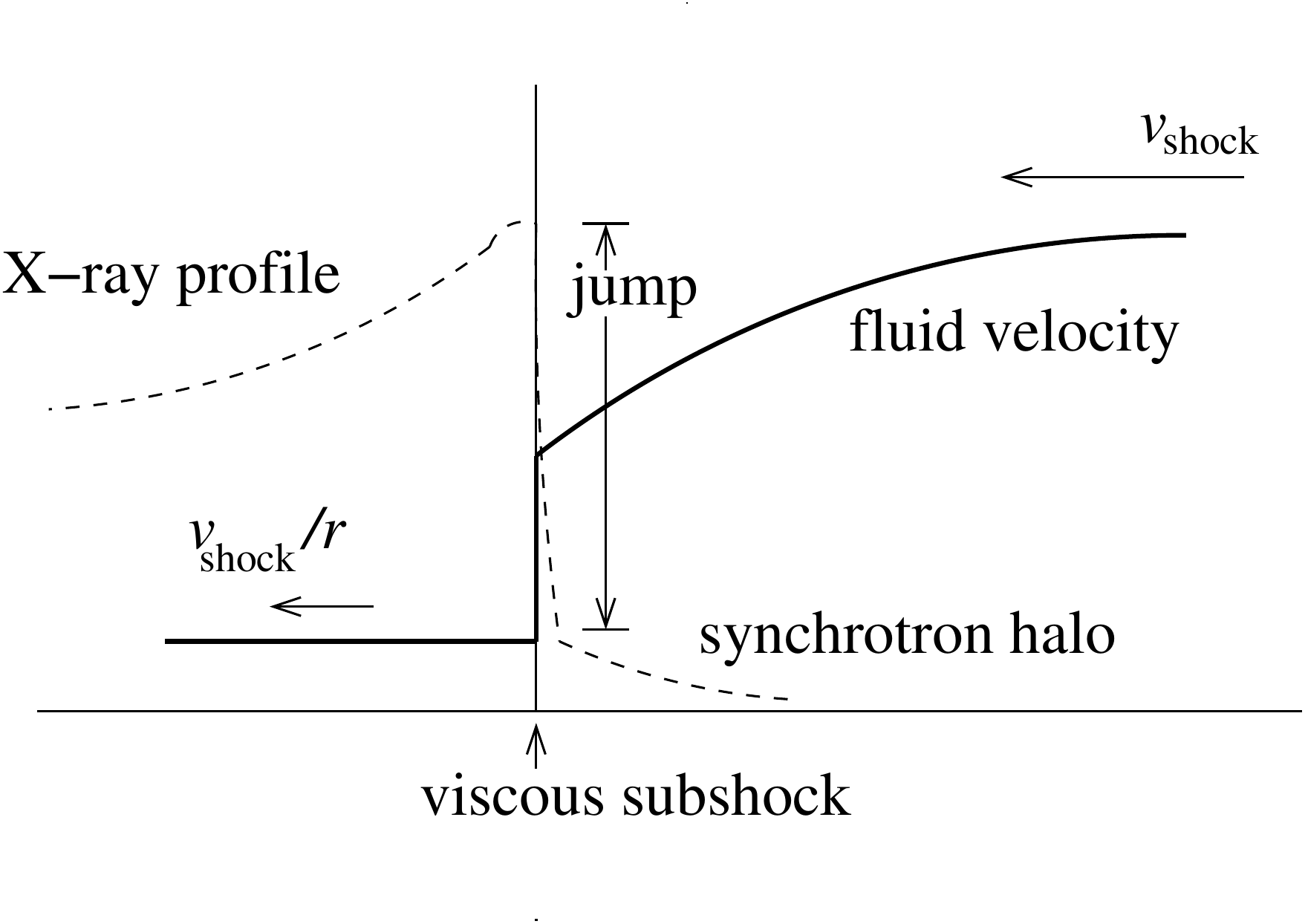}
\caption{This schematic illustrates the geometry (in the frame of the shock) for a cosmic-ray modified shock.  Gas flows in from the right with preshock velocity $v_{shock}$, and is gradually slowed before a sudden compression and deceleration at the viscous subshock.  The compressed gas then flows out to the left, with velocity $v_{shock}/r$, where $r$ is the compression ratio.  A precursor X-ray halo would extend into the region ahead of the viscous subshock where deceleration is taking place.
}
 \label{fig:halo_cartoon}
\end{figure}

\begin{figure}
\epsscale{0.90}
\plotone{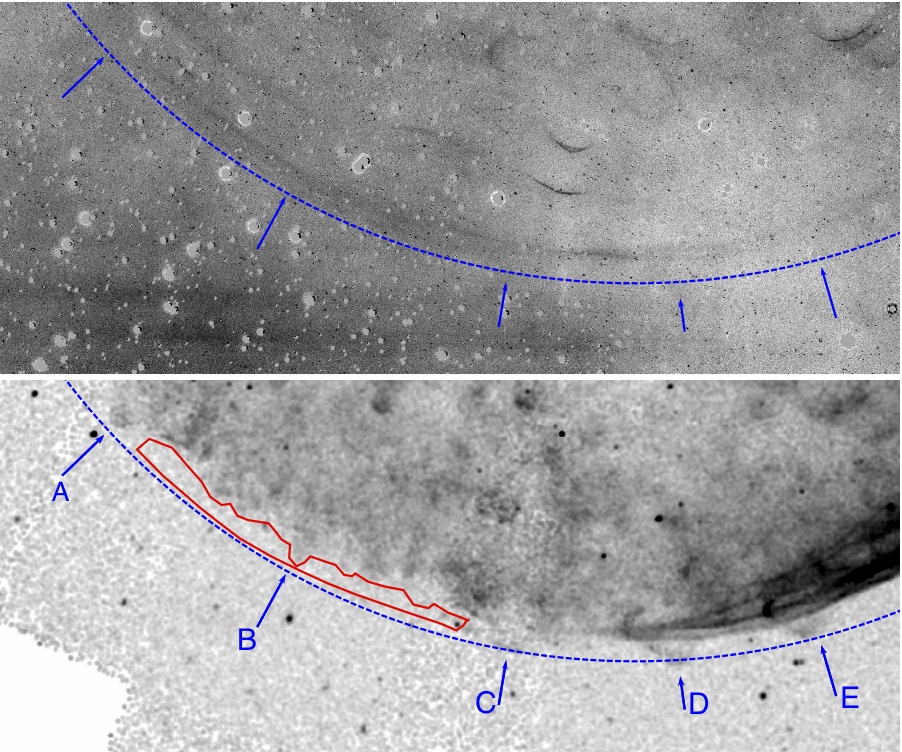}
\caption{The SE portion of the SN\,1006 in \ha\ (top) and broad X-rays (bottom), both displayed with a greyscale to show the faintest  emission.  The arrows indicate a number of clumps where thermal X-ray emission is found at or ahead of the shock as defined by the Balmer emission, which are probably concentrations of ejecta produced by instabilities.  The dashed circle \citep[with 15\farcm67 radius, centered at the same center from][which we use throughout this paper]{katsuda09} is simply to guide the eye and facilitate comparison.  In the sector from A to C, the outermost Balmer emission  is  $\sim 10\arcsec$ inside this circle, but farther west the outermost Balmer emission is farther inside.  Of the indicated X-ray clumps, A is located almost exactly at the Balmer front; B is well within, and C, D, and E are all well beyond it.  The region outlined in red, with width from 3\arcsec\ to 30\arcsec, was constructed to select an area {\em outside} the ejecta tufts, but {\em within} the outermost Balmer shell, in order to select for shocked ISM, as described in the text.
} 
\label{fig:south_protrusions}
\end{figure}

\end{document}

%% file: tab1.tex
\begin{deluxetable}{rcccccc}
\tablewidth{0pt}
\tablecaption{{\em Chandra} ACIS Observations of SN~1006}

\tablehead{
 \colhead{ObsID\tablenotemark{a}} &
 \colhead{Array} &
 \colhead{R.A. (J2000.)} &
 \colhead{Decl. (J2000.)} &
 \colhead{Roll} &
 \colhead{Obs.~Date} &
 \colhead{Exposure (ks)} 
}
\scriptsize
\tablewidth{0pt}\startdata
9107 &  ACIS-S & 15:03:51.5 &  -41:51:19 &  280.4\degr &  2008 Jun 24  &  89.0  \\
13737 & ACIS-S & 15:02:15.9 & -41:46:10 & \phn 31.7\degr  & 2012 Apr 20 & 87.1  \\
13738 & ACIS-I & 15:01:41.8 & -41:58:15 &  \phn 25.3\degr  & 2012 Apr 23 & 73.5  \\
14424 & ACIS-I & 15:01:41.8 & -41:58:15 &  \phn 25.3\degr  & 2012 Apr 27 & 25.4   \\
13739 & ACIS-I & 15:02:12.6 & -42:07:01 &  \phn  \phn  9.1\degr  & 2012 May 04 & 100.1\phn    \\
13740 &  ACIS-I &15:02:40.7 & -41:50:21 & 294.5\degr  & 2012 Jun 10 & 50.4  \\
13741 & ACIS-I & 15:03:48.0 & -42:02:53 &  \phn 24.6\degr  & 2012 Apr 25 & 98.5  \\
13742 & ACIS-I & 15:03:01.8 & -42:08:27 & 289.1\degr  & 2012 Jun 15 & 79.0  \\
13743 & ACIS-I & 15:03:01.8 & -41:43:05 &  \phn 19.9\degr  & 2012 Apr 28 & 92.6  \\
14423 & ACIS-I & 15:02:50.9 & -41:55:25 &  \phn 21.2\degr  & 2012 Apr 25 & 25.0  \\
14435 & ACIS-I & 15:03:42.5 & -41:54:49 & 297.3\degr  & 2012 Jun 08 & 38.3  
\enddata
\tablenotetext{a}{For ObsID 9107, Petre was PI; for the others, PI was Winkler.}
 \label{tab:acis_obsns}
 \end{deluxetable}

%% file: tab2.tex
\begin{deluxetable}{lccrr}
\tablewidth{0pt}
\tablecaption{CTIO Observations, 4m Blanco Telescope\tablenotemark{a}}

\tablehead{
\multicolumn{4}{c}{Filter} & 
 \\ 

\cline{1-4} 

\colhead{Informal} & 
\colhead{CTIO} &
\colhead{\phn$\lambda_c\tablenotemark{b}\;$} &
\colhead{$\Delta \lambda$\tablenotemark{b}} &
\colhead {\phn Exposure}   \\

\colhead{Name} &
\colhead{Designation} &
\colhead{(\AA)} &
\colhead{(\AA)} &
\colhead{(s)} 
}

\startdata

\ha & c6009 & 6563 &80\phn & \phn\phn\phn $24\times600$\phn\phn\phn   \\
\ha+8 nm & c6011 & 6650 & 80\phn &$ \phn\phn\phn 22\times600$\phn\phn\phn  \\

\enddata
\tablenotetext{a}{2010 April 15-18; observers: Winkler \& Long.}
\tablenotetext{b}{Central Wavelength and full width at half maximum in the f/2.8 telescope beam.}
 \label{tab:ctio_obsns}

\end{deluxetable}

%% file: tab3.tex
\begin{deluxetable}{lccc}
\tablecolumns{4}
\tablewidth{0pc}
\tabletypesize{\small}
\tablecaption{Equivalent Width Energy Bands} \tablehead{ 
\colhead{Element}  &
\colhead{Line Energy (keV)} & 
\colhead{Cont.\ (Low, keV)\tablenotemark{a}} & 
\colhead{Cont.\ (High, keV)\tablenotemark{a}}
}
\startdata

Oxygen & 0.51\,-\,0.74 &\phn  0.4\,-\,0.51 & 0.74\,-\,0.87\\
Neon & 0.89\,-\,0.97 & 0.74\,-\,0.87 & 1.12\,-\,1.2\phn\\
Magnesium & 1.29\,-\,1.42 & 1.20\,-\,1.29 & 1.42\,-\,1.7\phn\\
Silicon & \phn 1.7\,-\,1.95 & 1.42\,-\,1.7\phn & 1.95\,-\,2.2\phn\\

\enddata

\tablenotetext{a}{Cont.~(Low) and  Cont.~(High) give the continuum bands on
  either side of the line energy, used in producing the EQW images as described in the text.}
\label{eqwtable}
\end{deluxetable}

%% file: tab4.tex
\begin{deluxetable}{crcccccc}
\tablecolumns{8}
\tablewidth{0pc}
\tabletypesize{\small}

\tablecaption{Precursor Measurements} 
\tablehead{
\colhead{} & \colhead{} & \colhead{Post-Shock} & \colhead{Background\tablenotemark{a}} & \multicolumn{3}{c}{Net Pre-Shock Halo\tablenotemark{a,b}} & \colhead {Ratio\tablenotemark{c}} \\
\cline{5-7}

\colhead{Region} & \colhead{ObsID} & \colhead{Peak\tablenotemark{a}} &
 \colhead{20\arcsec --\,50\arcsec}& \colhead{ 0\arcsec --\,5\arcsec} & \colhead{ 5\arcsec --\,10\arcsec} & \colhead{ 10\arcsec --\,15\arcsec}  &  \colhead{Peak$/ $Halo\,(0\arcsec --\,5\arcsec)}
}

\startdata

W-2 & 9107 & 206  & $ \phn 4.5  \pm 0.6 $ &$ \phn 2.3 \pm 1.7  $ & $ \phn5.1  \pm 2.5$ & $ \phn 0.3 \pm 1.6$ &  \phn 89 \\
S-2 & 13739 & 312  & $\phn5.5  \pm 0.9 $ &$ \phn4.6  \pm 2.3$ & $ \phn3.8  \pm 3.2$ & $ - 1.3 \pm1.6$ &  \phn 68 \\
N-2 & 13743 & 125 & $\phn5.4  \pm 0.6 $ &$ \phn5.3  \pm 2.2$ & $ \phn3.2  \pm 2.2$ & $ \phn 0.6 \pm1.7$ &  \phn 23 \\
N-5 & 13743 & 390  & $\phn6.9  \pm 0.7 $ & $ 17.8  \pm 3.6$ & $ \phn2.6  \pm 2.6$ & $ \phn 2.8 \pm2.7$ &  \phn 22 \\
E-1 & 13738 & 461 & $12.5  \pm 1.0 $ & $ 24.4  \pm 4.9$ & $ \phn6.8  \pm 3.8$ & $ \phn 1.9 \pm 3.1$ &  \phn 19 \\
E-2 & 9107 & 463  & $12.4  \pm 0.9 $ & $ -0.1  \pm 2.3$ & $ \phn9.1  \pm 3.5$ & $ \phn 0.0 \pm 2.9$ &  \phn $>200$ \\

\enddata
\tablenotetext{a}{Surface Brightness ($10^{-10}$ photons cm$^{-2}$ s$^{-1}$ pixel$^{-1}$).}
\tablenotetext{b}{Measured surface brightness (after background subtraction) in indicated range ahead of shock.}
\tablenotetext{c}{Surface Brightness Ratio:  peak$/$net halo.}
\label{tab:halo_stats}
\end{deluxetable}  